\documentclass[pdflatex,sn-basic]{sn-jnl}
\usepackage{mathrsfs}
\usepackage{anyfontsize} 
\usepackage{lmodern}

\usepackage{graphicx}%
\usepackage{multirow}%
\usepackage{amsmath,amssymb,amsfonts}%
\usepackage{amsthm}%
\usepackage{mathrsfs}%
\usepackage[title]{appendix}%
\usepackage{xcolor}%
\usepackage{textcomp}%
\usepackage{manyfoot}%
\usepackage{booktabs}%
\usepackage{algorithm}%
\usepackage{algorithmicx}%
\usepackage{algpseudocode}%
\usepackage{listings}%

\usepackage[T1]{fontenc}
\usepackage[utf8]{inputenc} 
\usepackage{url}            
\usepackage{bm}             

\newcommand{\revv}[1]{\textcolor{black}{{#1}}}

\theoremstyle{thmstyleone}%
\theoremstyle{thmstyletwo}%

\theoremstyle{thmstylethree}%

\begin{document}
\title[The RAPID model]{Trajectory-Based Dust Evolution in Disks: First Results from the RAPID Simulation Code}


\author*[1,2]{\fnm{Dóra} \sur{Tarczay-Nehéz}}\email{tarczaynehez.dora@csfk.org}

\affil*[1]{\orgdiv{Konkoly Observatory}, \orgname{HUN-REN CSFK}, \orgaddress{\street{Konkoly Thege Miklós út 15-17}, \city{Budapest}, \postcode{1121}, \country{Hungary}}}

\affil[2]{\orgdiv{CSFK, MTA Centre of Excellence},  \orgaddress{\street{Konkoly Thege Miklós út 15-17}, \city{Budapest}, \postcode{1121}, \country{Hungary}}}


\abstract{The rapid depletion of dust particles in protoplanetary disks limits the time available for planetesimal formation, as solids are typically accreted onto the central star before dust particles can undergo substantial growth. Dust traps formed at sharp viscosity transitions — such as at the edges of the accretionally inactive dead zones — can halt radial drift and enhance dust coagulation. In this study, dust dynamics is investigated using \texttt{RAPID}, a one-dimensional Lagrangian-Eulerian simulation code that tracks representative particle trajectories over time. In order to explore the effect of physical parameters on dust evolution, a grid of 243 models was run. The simulation grid covers a range of parameters such as viscosity, width of the transition region at the edges of the dead zone, disk surface density exponent, and the collisional fragmentation velocity of the dust particles. The computational domain extends from 1 to 50 AU and covers $5\times10^5$ years of disk evolution, assuming a disk mass of $\sim 0.005\,M_\odot$. The results show that pressure maxima can trap up to $3-10\,M_\oplus$ of dust, depending on the local disk conditions. However, increasing the fragmentation velocity, decreasing the viscosity, or widening the dead zone transition width tends to reduce the effectiveness of dust trapping. The simulation results with \texttt{RAPID} reveal that dust evolution is highly sensitive to the physical conditions of the disk, which governs the early stages of planetesimal growth.}

\keywords{planet formation, protoplanetary disks, hydrodynamics, Lagrangian-Eulerian simulations}



\maketitle

\section{Introduction}

\label{sec:intro}

Planetesimals --- the building blocks of planets --- are thought to form through the accumulation and coagulation of dust grains in protoplanetary disks. 
One of the major bottlenecks in the core-accretion theory is the rapid loss of dust particles from the disk within a timescale that is much shorter than the disk's lifetime.
The timescale of the radial drift of the particles is dependent on their size and their radial distance from the star and the physical properties of the system (e.g., mass of the central star, gas surface density, etc).
According to \cite{Weidenschilling1980}, a $1$ meter-sized particle from $1$ AU can drift onto the central star on a timescale of $10^2-10^3$ years in an average system with a central solar-mass star.
As a result, most of the available dust is accreted onto the central star on timescales much shorter than those required for dust coagulation, thereby preventing the formation of larger aggregates.

The radial drift of the dust particles is caused by the difference between the orbital velocity of the sub-Keplerian gas and the Keplerian dust particles of the disk. 
This leads to a headwind that decelerates dust particles, thus they experience a continuous aerodynamic drag force (headwind) due to their motion relative to the sub-Keplerian gas \citep[see, e.g.,][and the references therein]{Binstiel2024}. 
The arising drag force causes the particles to steadily lose angular momentum, resulting in an inward spiral movement toward the central star.
An effective solution to overcome early dust depletion is based on the idea of a pressure maximum (often referred to as pressure or dust trap) in the disk \citep[e.g.,][]{PaardekooperandMellema2004,Riceetal2006}.
Here, the aerodynamic drag force effectively drops to zero, halting the radial drift of particles.

These traps often develop at places, where a steep viscosity gradient is present, such as at the inner and outer edges of the so-called dead zone, or at the edges of planet-carved gaps. Inside a pressure trap, dust particles can accumulate and their surface density increases. This enhances efficient coagulation of the particles due to their reduced relative collision velocities. 
Thus, these regions are considered to be promising regions for planetesimal formation and are possible planet nurseries \cite[see, e.g.,][]{Guileraetal2020}.

Several numerical studies have investigated such pressure traps in protoplanetary disks.
\cite{Rossby1939,Lovelaceetal1999} showed that at the vicinity of a sharp viscosity reduction \citep{Lyraetal2009b,Regalyetal2012}, complex, non-axisymmetric 2D (or 3D) structures, e.g., large-scale anticyclonic vortices, can be formed due to the excitation of the Rossby Wave Instability (RWI).
There phenomena act as an effective pressure trap in the protoplanetary disks \citep[e.g.,][]{BargeandSommeria1995,KlahrandHenning1997}.
The properties of these vortices --- such as their onset time, size, lifetime, azimuthal elongation --- are highly sensitive to disk parameters, including mass of the disk, viscosity of the gas, the self-gravity of the disk, and even the thermodynamical  properties, such as heating and cooling processes \cite[see, e.g.,][]{RegalyandVorobyov2017,PierensandLin2018,TarczayNehezetal2020,Tarczayetal2022}. 
Though these features require 2D or 3D modeling for a deeper understanding, after azimuthally averaging the radial profile of these structures, the asymmetries in the azimuthal direction smooth out, revealing a pressure maximum in the radial direction that can be well represented by 1D models.
For instance, a commonly used approximation in one-dimensional models is a stationary Gaussian gap-carved pressure bump \cite[see, e.g.,][]{Dullemondetal2018,Lauetal2022}.

In this work, a trajectory-based Lagrangian model, called \texttt{RAPID} (Representative Approach for Particle-Integrated Disks), is introduced. 
The model tracks the motion of the representative dust particles in the radial direction in an evolving 1D Eulerian gaseous disk model. 
A key difference in this approach is that the pressure maximum is not a static feature but a dynamically evolving phenomenon arising in the vicinity of the inner and outer edges of the embedded dead zone. 
This approach is computationally efficient due to its 1D nature, while following dust particles in a Lagrangian description, enabling effective approximation of particle trajectories. 
Utilizing \texttt{RAPID}, a parameter study is conducted to investigate dust mass growth, particle trajectories, and the evolution of feeding zones around pressure bumps, aiming to understand the conditions that facilitate planetesimal formation in protoplanetary disks.

\section{The \texttt{RAPID} model}
\label{sec:model}

To investigate the formation and viscous evolution of pressure maxima, a time-dependent one-dimensional hydrodynamic disk model (\texttt{RAPID}) is constructed. 
\texttt{RAPID} is designed to investigate the coupled evolution of the gas and dust material of the disk. 
It combines an Eulerian grid for the gas, and a Lagrangian description of the individual dust particles within a one-dimensional (radial) framework, assuming an axisymmetric disk.
This combination allows the user to make efficient investigations of various disk parameters and explore their effect on dust dynamics.

An earlier version of \texttt{RAPID} was used in \citep{Regalyetal2017} to model the evolution of dust particles in a dead zone edge (DZE) model. 
Since then, \texttt{RAPID} has undergone significant development. 
Coagulation and fragmentation of dust particles are also included, constrained by physical growth limits such as the fragmentation and the radial drift barriers \citep[see more details in e.g.,][]{Zsometal2008,Zsometal2010,Drkazkoskaetal2013,Cridlandetal2017}.
These processes are discussed in detail in Section~\ref{sec:particle_growth}. 
The implementation of these limiting effects is based primarily on the dust evolution model developed by \citet{Birnstieletal2012}, which captures the interplay between particle growth, fragmentation, and radial drift in turbulent protoplanetary disks.

Note, however, that the code is under active development.
By the time of the submission, it is not publicly available, as it is undergoing further refinement, but it will be published on GitHub\footnote{https://github.com/tnehezd/RAPID} in the near future.

\subsection{Unit system}

\texttt{RAPID} uses a unit system commonly adopted in protoplanetary disk modeling: distances are measured in astronomical units (AU), masses in solar masses ($M_\odot$), and time in years. In this system, the gravitational constant is dimensionless and set to $G = 1$, which both simplifies the equations of motion and ensures that a particle orbiting at 1\,AU around a 1\,$M_\odot$ star has an orbital period of $2\pi$ years.

\subsection{Gaseous disk hydrodynamics}
\label{sec:gashydro}
The evolution of the gaseous material in the disk is modeled as a geometrically thin, axisymmetric, and viscously evolving accretion disk \citep{Pringle1981, Lodato2008, Armitage2010,Armitage2011}.
Assuming an axisymmetric and vertically thin disk, the full Navier -- Stokes equations can be vertically integrated over the disk thickness, reducing the problem to a one-dimensional description in the radial direction.
This yields an advection-diffusion type equation governing the time evolution of the gas surface density, $\Sigma_g$, which is initially set to a power-law profile in the form of $\Sigma_\mathrm{g} \propto r^{-p}$, and includes the effects of viscous angular momentum transport in the disk.

Following the classical treatment by \citet{LyndenBellandPringle1974}, the evolution of the surface density is described by:
\begin{equation}
\frac{\partial \Sigma_\mathrm{g} }{\partial t} = \frac{3}{r} \frac{\partial}{\partial r} \left[ r^{1/2} \frac{\partial}{\partial r} \left(\nu \Sigma_\mathrm{g} r^{1/2}\right) \right], 
\label{eq:gas_continuity_general}
\end{equation}

\noindent where $\nu$ is the kinematic viscosity of the gas, and $r$ is the radial distance from the star.
The kinematic viscosity, $\nu$, is parameterized using the well-known $\alpha$-prescription of \citep{ShakuraandSunyaev1973}:

\begin{equation}
\nu = \alpha H c_\mathrm{s},
\label{eq:viscosity}
\end{equation}

\noindent where $\alpha  \in \left [ 0 , 1 \right]$ is the dimensionless parameter, that accounts for the angular momentum transport in the disk. 
Observations suggest that $\alpha$ typically lies in the range $10^{-3}$–$10^{-2}$.
In equation \ref{eq:viscosity}, $H = hr$ describes the pressure scale height of the disk measured from the mid-plane (with a geometric aspect ratio of $h$), and $c_\mathrm{s}$ is the local sound speed.

In the case of a locally isothermal disk, $c_\mathrm{s}$ can be expressed as:

\begin{equation}
    \label{eq:soundspeed}
    c_\mathrm{s} = H \Omega_\mathrm{K},
\end{equation}

\noindent where $\Omega_K$ is the Keplerian angular velocity of the gas around an $M_\star$ mass star (with $G$ being the dimensionless universal gravitational constant):

\begin{equation}
    \label{eq:omega}
    \Omega_\mathrm{K} = \sqrt{\frac{G M_\star}{r^3}}.
\end{equation}

\noindent This way, $\Omega_\mathrm{K}$ is connected to the Keplerian velocity of the gas at a given $r$ radius:

\begin{equation}
    \label{eq:kep_vel}
    v_\mathrm{K} = r \Omega_\mathrm{K}.
\end{equation}

As a locally isothermal disk model is assumed, the equation of state of the gas is expressed as

\begin{equation}
    P = \Sigma_\mathrm{g} c_\mathrm{s}^2,
\end{equation}

\noindent where $P$ is the vertically integrated pressure at a given radius in the midplane.

\begin{figure}
    \centering
    \includegraphics[width=\linewidth]{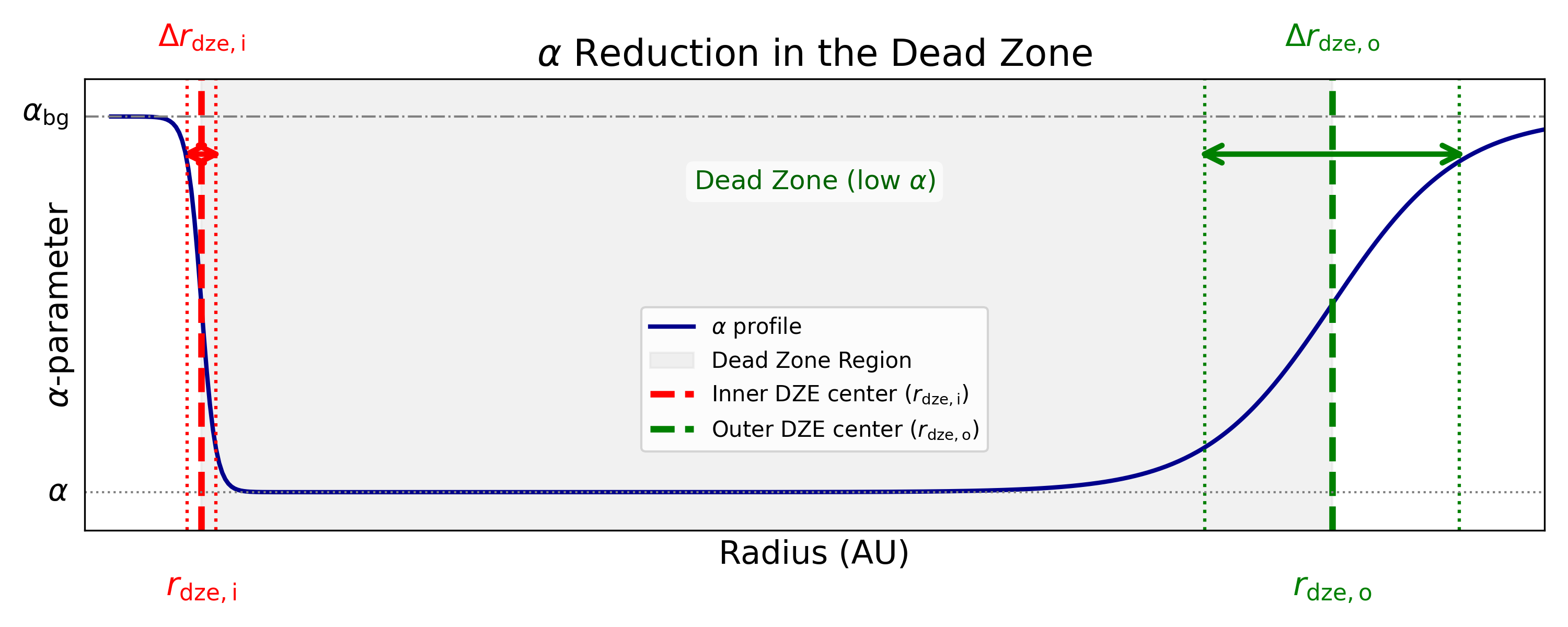}
    \caption{A scematic view of the reduction of the $\alpha$ parameter described in Equation\,\ref{eq:alphared}. Here, the adopted values for the inner ($r_\mathrm{dze,i}$) and outer ($r_\mathrm{dze,o}$) dead zone edges are 2.7 and 24 AU, respectively. The width of the dead zone edge is defined by $\Delta r_\mathrm{dze,i/o} = 2.0\,H$. As $\Delta r_\mathrm{dze,i/o}$ depends on the local pressure scale height ($H$), the radial extent in AU differs between the two edges.}
    \label{fig:alpha_red}
\end{figure}

A main feature of \texttt{RAPID} is the explicit implementation of the dead zone of the disk \citep{Gammie1996}. 
These low-ionization (accretionally “inactive”, or “dead”) regions are characterized by suppressed magneto-rotational instability (MRI), resulting in significantly reduced turbulent viscosity.
In the \texttt{RAPID} model, following the idea of \cite{Lyraetal2009b}, dead zone is modeled with reducing the kinematic viscosity of the gas by a factor, $\delta_\alpha$, as

\begin{equation}
\label{eq:alpha}
    \alpha = \delta_\alpha \cdot \alpha_\mathrm.
\end{equation}

\noindent Here $\alpha$ represents the value of the \cite{ShakuraandSunyaev1973} $\alpha$ parameter throughout the disk outside the dead zone. 
The reduction of the viscosity is described by a combined $\tanh$ function \cite[see, e.g.,][]{Regalyetal2012} as follows:

\begin{equation}
\label{eq:alphared}
\delta_{\alpha} = 1 - \frac{1}{2}\left ( 1 - \alpha_{\mathrm{mod}}  \right ) \left [  \tanh{\left ( \frac{r - r_{\mathrm{dze,i}}}{\Delta r_{\mathrm{dze,i}}} \right ) }+  \tanh{\left ( \frac{r_{\mathrm{dze,o}} - r}{\Delta r_{\mathrm{dze,o}}} \right )} \right].
\end{equation}

\noindent Here, $\alpha_{\mathrm{mod}}$ is the depth of the reduction, $r_{\mathrm{dze,i}}$, $r_{\mathrm{dze,o}}$ and $\Delta r_{\mathrm{dze,i}}$, $\Delta r_{\mathrm{dze,o}}$ are the location and the half-width of the viscosity reduction at the inner (i) and outer (o) edges of the accretionally inactive dead zone.
Figure\,\ref{fig:alpha_red} illustrates the profile of $\alpha$ reduction in the vicinity of the dead zone ($r_\mathrm{dze,i} = 2.7$ AU, $r_\mathrm{dze,o} = 24$ AU, $\Delta r_\mathrm{dze,i/o} = 2\,H$).
As the viscosity reduction, $\Delta r_\mathrm{dze,i/o}$, depends on the local pressure scale height, the actual radial extent in AU differs between the inner and outer edges of the dead zone.

Note that throughout this paper, $\Delta r_\mathrm{dze,i/o}$ is expressed in units of the local pressure scale height.
Thus, the width of the dead zone edges is a function of the distance, rather than a fixed, arbitrary set value.
This normalization allows us for a direct comparison between the inner and outer dead zone edges.

The \texttt{RAPID} code solves Equation\,\ref{eq:gas_continuity_general} numerically using a finite-difference scheme on an equidistant radial grid.
The right-hand side is discretized explicitly by computing central differences of the quantity $\nu \Sigma_\mathrm{g}$, and the surface density is updated via a forward Euler step.
The timestep is constrained by the Courant–Friedrichs–Lewy (CFL) condition to ensure numerical stability. This restriction prevents information from propagating beyond one grid cell per step, thereby maintaining the physical consistency of the solution.

\subsection{Dust Particle Dynamics and Size Evolution}
\label{sec:particle_growth}

In \texttt{RAPID}, the dusty material of the protoplanetary disk is modeled by using a representative particle approach.
This method employs an ensemble of $N_\mathrm{p}$ Lagrangian particles, enabling detailed tracking of individual particle properties and trajectories.
The approach discretizes the disk into $N_\mathrm{p}$ Lagrangian dust particles, where $N_p$ does not necessarily match the resolution $N_\mathrm{grid}$ of the gas grid.
Each particle represents the dust mass within a specific annulus (or cell in 1D) at its initial radial position, which remains fixed throughout the simulation.
Furthermore, each particle is assigned a unique index (ID) at initialization, which remains fixed throughout the simulation and serves to identify the particle across all timesteps.
Thus, every particle is uniquely characterized by its initial ID, mass (and associated surface density), and its current position at each timestep.
This setup facilitates straightforward monitoring of dust transport and accumulation within pressure maxima.
Since the dust is treated as discrete particles rather than a continuous fluid, their trajectories can be precisely followed --- a feature particularly useful for identifying the “feeding zones” near the edges of dead zones, i.e. the radial regions around pressure maxima that collect dust particles.

Furthermore, \texttt{RAPID} is designed to simultaneously track two distinct dust grain populations in the same simulation: a primary population of centimeter-sized particles and a background population of micrometer-sized grains.

\subsubsection{Radial drift barrier}
\label{sec:drift_barrier}

As described in Section~\ref{sec:intro}, the gaseous component of the disk orbits the central star with a sub-Keplerian velocity ($v_{\phi,\mathrm{gas}} < v_\mathrm{k}$), while dust particles tend to move with the local Keplerian velocity.
This velocity difference produces a persistent headwind on the dust particles, causing continuous angular momentum loss via drag forces.
Consequently, the dust gradually migrates inward, leading to a depletion of solid material in the disk.

The motion of dust particles is governed in Cartesian coordinates by the equation:

\begin{equation}
\label{eq:eqofmotion}
\frac{\mathrm{d}^2\mathbf{r}}{\mathrm{d}t^2} = - G \frac{M_\star}{\left | \mathbf{r} \right |^3} \mathbf{r} + \mathbf{f_\mathrm{D}},
\end{equation}

\noindent where $\mathbf{r}$ is the particle’s position vector, $M_\star$ is the mass of the central star, and $G=1$ is the dimensionless universal gravitational constant.
\texttt{RAPID} solves Equation\,\ref{eq:eqofmotion} in one dimension along the radial coordinate in polar geometry, assuming axisymmetry, using a classical fourth-order Runge–Kutta scheme (RK4).
At each RK4 stage, the drag force is calculated explicitly using gas properties interpolated from the disk, and the particle position is updated accordingly.

The acceleration due to drag, following \cite{Lyraetal2009b}, is given by:
\begin{equation}
\label{eq:drag_force}
\mathbf{f_\mathrm{D}} = - \left(\frac{3 \rho_0 C_\mathrm{D} \left| \Delta \mathbf{v} \right |}{8 \rho_\mathrm{p} s_\mathrm{p}}\right) \Delta \mathbf{v},
\end{equation}
\noindent where $\rho_0 = (1/\sqrt{2\pi})\cdot \Sigma/H$ is the midplane gas density, $\rho_\mathrm{p}$ and $s_\mathrm{p}$ are the intrinsic density and size of the dust grain, and $\Delta \mathbf{v} = \mathbf{u} - \mathbf{v_\mathrm{g}}$ is the relative velocity between the dust ($\mathbf{u}$) and gas ($\mathbf{v_\mathrm{g}}$).
The coefficient $C_\mathrm{D}$ is a dimensionless friction factor that smoothly interpolates between the Epstein ($s \lesssim \lambda$) and Stokes ($s \gtrsim \lambda$) drag regimes \cite[see, e.g.,][]{WoitkeandHelling2003,Paardekooper2007}.
Here, $\lambda$ is the mean free path of gas molecules, quantifying the average distance a molecule travels before colliding.

The coupling strength between dust and gas is described by the dimensionless Stokes number, $\mathrm{St}$, defined as the ratio of the particle’s frictional timescale ($t_\mathrm{fric}$) to the dynamical (orbital) timescale ($t_\mathrm{D} = \Omega^{-1}$):
\begin{equation}
\label{eq:II_Stokes_midplane}
\mathrm{St} = \frac{s\rho_{\mathrm{p}}}{\Sigma_{\mathrm{g}}}\frac{\pi}{2}.
\end{equation}
\noindent Particles with $\mathrm{St} \ll 1$ are tightly coupled to the gas and quickly adjust to its sub-Keplerian velocity. 
In this regime, the gas continuously forces the particles to orbit slower than their Keplerian speed, resulting in a steady loss of angular momentum and an inward drift.
Naturally, as $\mathrm{St}$ decreases, the radial drift of the particles slows down to the point where its timescale becomes comparable to the viscous evolution timescale of the gas.

In contrast, particles with $\mathrm{St} \gg 1$ are weakly coupled and tend to orbit near the Keplerian velocity.
As a result, they experience a drag force from the gas, leading to a slower but persistent radial drift, as well.
This is expressed in the followings.
In one dimension, according to \cite[e.g.,][]{Nakawagaetal1986, Youdinetal2007, TakeuchiandLin2002, Birnstieletal2012}, the radial velocity of a dust particle ($u_\mathrm{dust,r}$) reads as follows,

\begin{equation}
\label{eq:dustrad}
    u_\mathrm{dust,r} = \frac{u_\mathrm{gas,r}}{1+\mathrm{St}^2} + \frac{2}{\mathrm{St} +\mathrm{St}^{-1}}u_\mathrm{drift},
\end{equation}

\noindent where $u_\mathrm{gas,r}$ is the radial velocity of the gas:

\begin{equation}
\label{eq:ugas}
    u_\mathrm{gas,r} = - \frac{3}{\Sigma_\mathrm{g} r^{1/2}}\frac{\partial}{\partial r}\left( \Sigma_\mathrm{g} \nu r^{1/2}\right),
\end{equation}

\noindent and  $u_\mathrm{drift}$ is the maximum drift velocity of a particle:

\begin{equation}
\label{eq:udrift}
    u_\mathrm{drift} = \frac{c_\mathrm{s}^2}{2\Omega_\mathrm{K} r}\frac{d\ln P}{d \ln r}.
\end{equation}

\noindent From Equation\,\ref{eq:dustrad}, it can be seen, that the fastest radial drift occurs for particles with $\mathrm{St} \sim 1$, where the difference between gas and dust velocity reaches its maximum value.

The timescale of the radial drift can be expressed as

\begin{equation}
\label{eq:taudrift}
    \tau_\mathrm{drift} = \frac{r}{\left| u_\mathrm{drift}\right|}.
\end{equation}

\noindent Substituting Equation \ref{eq:udrift}, \ref{eq:soundspeed} and \ref{eq:kep_vel}
into \ref{eq:taudrift} the drift timescale can also be expressed as

\begin{equation}
\label{eq:tau_drift_alt}
\tau_\mathrm{drift} = \frac{r v_\mathrm{k}}{\mathrm{St} c_\mathrm{s}^2} \gamma^{-1},
\end{equation}

\noindent where $\gamma$ is the value of the logarithmic pressure gradient:

\begin{equation}
    \gamma = \left| \frac{\mathrm{d} \ln P}{\mathrm{d} \ln r} \right|.
    \label{eq:gammadrdp}
\end{equation}

\noindent 

Introducing $\epsilon = \Sigma_\mathrm{d} / \Sigma_\mathrm{g}$ as the dust-to-gas ratio, and 

\begin{equation}
\label{eq:growthts}
 \tau_\mathrm{gr} \simeq \frac{1}{\epsilon\Omega_\mathrm{K}}
\end{equation}

\noindent as the growth timescale of the particles, the Stokes number of the maximum size of the dust particles limited by radial drift, $\mathrm{St_{drift}}$, can be expressed by equating $\tau_\mathrm{gr}$ and $\tau_\mathrm{drift}$:

\begin{equation}
\label{eq:Stdrift}
    \mathrm{St_{drift}} = \frac{r v_\mathrm{K} \epsilon \Omega_\mathrm{K}}{c_\mathrm{s}^2} \gamma^{-1}.
\end{equation}

\noindent Substituting Equation \ref{eq:kep_vel} into \ref{eq:Stdrift} the latter can be expressed as

\begin{equation}
\label{eq:Stdrift_fd}
\mathrm{St}_\mathrm{drift} = f_\mathrm{drift} \epsilon \frac{v_\mathrm{K}^2}{c_\mathrm{s}^2} \gamma^{-1},
\end{equation}

\noindent where $f_\mathrm{d}$ is a dimensionless factor that accounts for specific model assumptions and numerical coefficients \cite[see more details in][]{Birnstieletal2012}.

As a result, assuming an Epstein regime, the drift limited particle size can be calculated as \cite{Birnstieletal2012}:

\begin{equation}
    \label{eq:adrift}
    a_\mathrm{drift} = f_\mathrm{drift} \frac{2}{\pi}\frac{\Sigma_\mathrm{d}}{\rho_\mathrm{p}}\frac{v_\mathrm{K}}{c_\mathrm{s}^2}\gamma^{-1}.
\end{equation}

\subsubsection{Turbulent fragmentation barrier}
\label{sec:frag_barrier}

Considering dust coagulation, fragmentation due to background turbulence hinders dust particles with $\mathrm{St} \ll 1$ from growing in size beyond a certain limit, what is set by the fragmentation threshold velocity, $u_\mathrm{frag}$ \citep[see, e.g.,][]{Braueretal2008}. 
The fragmentation threshold velocity strongly depends on the physical properties of the dust grains. 
For example,
laboratory experiments have revealed that the threshold velocities are around 1 m/s for silicate dust grains \citep[see, e.g.,][]{BlumandWurm2008,Wadaetal2008, Gundlachetal2011}, and $\sim 50$ m/s \citep{Wadaetal2009} for ice particles. 

Similarly to the drift limited case, the fragmentation barrier can be expressed in terms of a limiting  Stokes number, $\mathrm{St_{frag}}$, \citep[see, e.g.,][]{Birnstieletal2009}:

\begin{equation}
\label{eq:St_frag}
\mathrm{St}_\mathrm{frag} = \frac{1}{\alpha} \left(\frac{u_\mathrm{frag}}{c_\mathrm{s}}\right)^2,
\end{equation}

\noindent where $\alpha$ is calculated in Equations \ref{eq:alpha} and \ref{eq:alphared} and $c_\mathrm{s}$ is the local isothermal sound speed of the gas, see Equation \ref{eq:soundspeed}. 

According to \cite{Birnstieletal2012}, a significant fraction of the dust particles' size is slightly below the limiting Stokes number, $\mathrm{St}_\mathrm{frag}$.
Hence, a dimensionless parameter, $f_\mathrm{frag}$ is applied to calculate the size limit, $a_\mathrm{frag}$:

\begin{equation}
    \label{eq:afrag}
    a_\mathrm{frag} = f_\mathrm{frag} \frac{2}{3\pi} \frac{\Sigma_\mathrm{g}}{\rho_\mathrm{p}} \alpha \frac{u_\mathrm{frag}^2}{c_\mathrm{s}^2}.
\end{equation}

\subsubsection{Drift-induced fragmentation barrier}
\label{sec:relvelfrag}

Drift-induced fragmentation barrier accounts for fragmentation caused by relative velocities of the particles due to radial drift.
This barrier is relevant in very low turbulent regimes, like the dead zone of the disk.
To obtain the size limit for this barrier, the relative velocity of the colliding particles with Stokes numbers $\mathrm{St_1}$ (bigger) and  $\mathrm{St_2}$ (smaller), $u_\mathrm{df}$ is expressed as

\begin{equation}
    \label{eq:relveldrift}
    \Delta u_\mathrm{df} = \frac{c_\mathrm{s}^2}{v_\mathrm{K}} \gamma \left| \mathrm{St_1} -  \mathrm{St_2}\right|,
\end{equation}

\noindent which is valid only in the limit $\mathrm{St} \ll 1$ \citep{Birnstieletal2012}. 
The Stokes number of the smaller particle can be expressed as a function of the bigger one as:

\begin{equation}
    \label{eq:st2}
    \mathrm{St_2} = k \mathrm{St_1},
\end{equation}

\noindent where $k$ is a dimensionless parameter that describes the efficiency of fragmentation during collisions, allowing for the determination of a size limit \citep{Birnstieletal2012}.
Fragmentation of larger particles is the most efficient when the collision takes place with another similar sized particle, thus $k=0.5$ is applied in \texttt{RAPID}.

This way, the limit of the dust-induced fragmentation barrier can be expressed with the Stokes number, $\mathrm{St_{df}}$ as follows:

\begin{equation}
    \label{eq:Stdf}
    \mathrm{St_{df}} = \frac{u_\mathrm{frag} v_\mathrm{k}}{\gamma c_\mathrm{s}^2 (1-k)},
\end{equation}

\noindent and the size limit of the particle is

\begin{equation}
    a_\mathrm{df} = u_\mathrm{frag}\frac{v_\mathrm{K}}{\gamma c_\mathrm{s}^2 (1-k)} \frac{2\Sigma_\mathrm{g}}{\pi\rho_\mathrm{p}}.
\end{equation}

\subsection{Particle size evolution}

\texttt{RAPID} calculates the evolution of particle's size following the idea of a simple model of \cite{Birnstieletal2012}. Based on Equation~\ref{eq:growthts}, the characteristic time required for a particle to grow from an initial size $a_0$ to a size of $a_1$ can be estimated by

\begin{equation}
    t_\mathrm{gr} = \tau_\mathrm{gr} \ln \left( \frac{a_1}{a_0}\right).
\end{equation}

\noindent Here, $a_1$ can be determined as the limiting particle size set by the above-mentioned barriers (see Sections\,\ref{sec:drift_barrier}, \ref{sec:frag_barrier} and \ref{sec:relvelfrag}):

\begin{equation}
a_1 = \min(a_{\mathrm{frag}}, a_{\mathrm{drift}},a_\mathrm{df}).
\label{eq:sizelimit}
\end{equation}

\noindent Although Equation\,\ref{eq:sizelimit} defines the theoretical upper size limit for particle growth, this value is not reached instantaneously. Instead, the particle size evolution is governed by the local growth timescale $\tau_{\mathrm{grow}}$, which depends on disk properties and the dust-to-gas ratio.
The theoretical particle size resulting from exponential growth can be expressed as

\begin{equation}
\tilde{a} = a_0 \cdot \exp\left(\frac{t-t_0}{\tau_{\mathrm{grow}}}\right),
\end{equation}

\noindent here $t_0$ is the initial $t=0$ state.
The actual particle size at time $t$  is therefore determined by the competition between the physical size limit $a_1$ and the model-predicted growth $\tilde{a}$, resulting in

\begin{equation}
a = \min\left(a_1,\ \tilde{a}\right).
\end{equation}

\noindent This ensures that particle growth never exceeds the local size barriers.

Upon calculating the actual size of a particle, its Stokes number (see Equation\,\ref{eq:II_Stokes_midplane}), that describes the coupling to the gas, can be determined as

\begin{equation}
\label{eq:stokes_actual}
    \mathrm{St} = \frac{a \rho_\mathrm{p}}{\Sigma_\mathrm{g}}\frac{\pi}{2}.
\end{equation}

\noindent Substituting Equation\,\ref{eq:stokes_actual} into \ref{eq:dustrad}, the radial velocity ($u_\mathrm{dust,r}$) of each particles can be calculated.

\section{Simulation Setup}
\label{sec:simsetup}

To investigate the effect of physical properties of the protoplanetary disk, a grid was constructed utilizing the \texttt{RAPID} code. 
As detailed previously in Section \ref{sec:model}, this is a one-dimensional (1D), vertically integrated model that tracks the time-dependent behavior of the gas disk and the embedded dust particles. 
It is important, however, to note that the current version of \texttt{RAPID} handles only a one-way coupling between the gas and dust material, meaning the dust evolves within the gas disk, but the dust does not exert a back-reaction on the gas.

\begin{table}[tbp]
\caption{Used parameters for the simulation grid.}
\label{tab:setup}
\begin{tabular}{@{}llll@{}}
\toprule
    \textbf{Parameter} & \textbf{Symbol} & \textbf{Values Explored} & \textbf{Unit} \\
\midrule
\midrule
    \multicolumn{4}{l}{\textbf{Disk Properties}} \\
\midrule
     $\alpha$ Parameter & $\alpha$ & 0.01, 0.001, 0.0001 & - \\
    $\alpha$ Reduction & $\alpha_\mathrm{mod}$ & 0.01, 0.001, 0.0001 & -\\
    Gas Surface Density\footnotemark[1] at $r=1$ & $\Sigma_0$ & $29.6,\ 142.2,\ 573.5$ & $\mathrm{g/cm^2}$\\
    Inner Disk Radius & $R_\mathrm{in}$ & 1 & AU \\
    Outer Disk Radius & $R_\mathrm{out}$ & 50 & AU \\
    Disk Surface Density Exponent & $p$ & 0.5, 1.0, 1.5 & - \\
    \hline
    \multicolumn{4}{l}{\textbf{Dust Properties}} \\
    \hline
    Critical Fragmentation Velocity & $u_\mathrm{frag}$ & 100, 300, 500 & cm/s \\
    Particle Density & $\rho_\mathrm{p}$ & 1.6 & g/cm$^3$ \\
    Size of the Dust Particles & $s_\mathrm{p}$ & 1 & cm\\
    \hline
    \multicolumn{4}{l}{\textbf{Simulation Features}} \\
    \hline
    Two-Population Model & - & Disabled & - \\
    Dust Size Evolution & - & Enabled & - \\
    Dust Radial Drift & - & Enabled & - \\
    Gas Disk Evolution & - & Enabled & - \\
    \hline
    \multicolumn{4}{l}{\textbf{Numerical Resolution}} \\
    \hline
    Number of Grid Cells & $N_\mathrm{grid}$ & 1000 & - \\
    Number of Dust Particles & $N_\mathrm{p}$ & 1000 & - \\
    \hline
    \multicolumn{4}{l}{\textbf{Dust Accumulation Zones}} \\
    \hline
    Inner Zone Radius & $r_\mathrm{dze,i}$ & 2.7 & AU \\
    Outer Zone Radius & $r_\mathrm{dze,o}$ & 24 & AU \\
    DZE Width Factor\footnotemark[2] & $\Delta r_\mathrm{dze,i/o}$ & 1.0, 1.5, 2.0 & $H$ \\
\botrule
\end{tabular}
\footnotetext[1]{In \texttt{RAPID}, surface densities are expressed in $M_\odot/\mathrm{AU}^2$. 
For consistency with common usage, these values were converted to $\mathrm{g/cm^2}$ in this paper, using 
$1\ M_\odot/\mathrm{AU}^2 \approx 8.89\times10^{6}\ \mathrm{g/cm^2}$.}
\footnotetext[2]{$\Delta r_\mathrm{dze,i/o}$ is measured in local pressure scale heights, see more in Section \ref{sec:gashydro}}

\end{table}

Table \ref{tab:setup} summarizes the utilized parameters for the simulation. 
This yields altogether a set of 243 different types of parameter setups.
All simulations include dust particle's growth (via dust coagulation and fragmentation discussed in Section\,\ref{sec:particle_growth}), dust radial drift, and gas disk evolution over time, and each run was performed for $5\times 10^5$ years. 
The two-population feature of \texttt{RAPID}, however, was excluded to facilitate faster grid investigation, focusing on a single dust population.

\subsection{Disk Geometry and Physical Properties of Gas}
\label{sec:diskgeom}

The computational domain extends radially from an inner boundary at $R_\mathrm{in} = 1 \, \mathrm{AU}$ to an outer boundary at $R_\mathrm{out} = 50 \, \mathrm{AU}$ in each disk setup within the computational grid. 
The radial grid consists of $N_\mathrm{grid} = 1000$ equidistant cells, providing a uniform linear resolution in radial distance.
As discussed earlier, the viscosity of the disk is modeled by the $\alpha$-prescription of \cite{ShakuraandSunyaev1973}, assuming a flat ($H = hr$), 
locally isothermal disk.
The geometric aspect ratio of the disk is assumed to be $h=0.05$ uniformly for the whole computational grid.

As discussed in Section \ref{sec:gashydro}, the surface density of the gas, $\Sigma_\mathrm{g}$, follows a power-law profile as

\begin{equation}
    \Sigma_\mathrm{g} = \Sigma_0 r\left(\frac{r}{r_0}\right)^{-p}.
\label{eq:sig_profile}
\end{equation}

\noindent Here $\Sigma_0$ is the value of the surface density of the gas at a given reference value ($r_0=1$ AU in these models \footnote{In \texttt{RAPID}, surface densities are expressed in $M_\odot/\mathrm{AU}^2$. For the sake of consistency, these values are converted to $g/cm^2$ with the conversion rate of 
$1\,M_\odot/\mathrm{AU}^2 \approx 8.89\times10^{6}\,\mathrm{g/cm^2}$.}). 

The adopted $\Sigma_0$ values are $29.6$ $\mathrm{g/cm^2}$, $142.2$  $\mathrm{g/cm^2}$, and $573.5$  $\mathrm{g/cm^2}$ for $p=0.5$, $1.0$, and $1.5$ models, respectively.

\begin{figure}
    \centering
    \includegraphics[width=\linewidth]{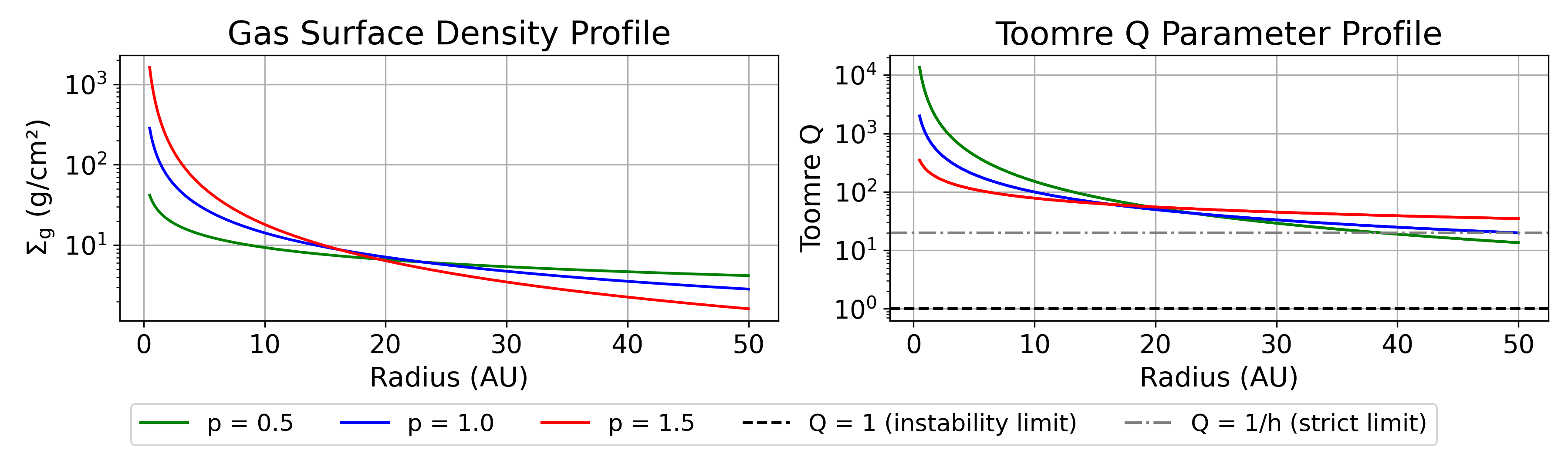}
    \caption{Initial gas surface density profile for $p=0.5$, $1$ and $1.5$ models (right panel) and the corresponding initial Toomre $Q$ parameter for each cases. Dashed lines present the $Q=1$ stability limit, while the dashed-dotted line corresponds to the strict $Q_\mathrm{crit} = 1/h$ stability limit.}
    \label{fig:toomre}
\end{figure}

$\Sigma_0$ values were chosen that the total gas disk mass between $1$ and $50$ AU is $\sim 0.005\,M_\odot$ for all cases. 
Thus, the integrated disk mass is a factor of 2 lower than the typical $0.01\,M_\odot$ associated with the Minimum Mass Solar Nebula (MMSN) for gas \cite{Weidenschillingetal1977b,Hayashi1981}. 
This choice helps to ensure that initially, the disk is stable against its self-gravity, that is parametrized by the Toomre parameter $Q$ as follows:

\begin{equation}
    Q = \frac{c_\mathrm{s} \Omega_\mathrm{K}}{\pi G \Sigma_\mathrm{g}}
\end{equation}

Commonly, a protoplanetary disk is considered to be unstable, if  $Q < 1$. 
Figure\,\ref{fig:toomre} shows that in all cases, the $Q$ parameter remains above $1$, fulfilling the criterion of a stable disk.
Note, however, that theoretical work of \cite{LovelaceandHohlfeld2013} and \cite{Yellinetal2016} revealed that the effect of self-gravity is not negligible if $Q < Q_\mathrm{crit}$, where $Q_\mathrm{crit} = 1/h$. 
This yields to a threshold of $20$ in this paper. Thus, in the case of $p=0.5$, $Q$ drops down below $Q_\mathrm{crit}$ at around $38.5$ AU, while for $p=1$ models, $Q$ approaches $Q_\mathrm{crit}$ at $\sim50$ AU.
In both cases, these regions are beyond the outer boundaries of the dead zone, which lays in the primary focus of this study.
Reducing the total disk mass further to fulfill stability even in these outermost regions leads to a disk mass that differs more from the typical values suggested by MMSN model (i.e., $\sim 0.003\,M_\odot$ for $p=0.5$ and $\sim 0.0004\,M_\odot$ for $p=1.0$).

The effect of self-gravity has been studied in two-dimensional simulations, showing that it can significantly influence the lifetime, size, and number of emerging vortices \cite[see, e.g.,][and the references therein]{RegalyandVorobyov2017,TarczayNehezetal2020,Tarczayetal2022}. 
While such features cannot be directly captured in our one-dimensional framework, the vortex lifetime can still be investigated, suggesting that considering self-gravity may be relevant. 
However, in this study, self-gravity was not included in the simulations, as this effect is not implemented in RAPID.

Therefore, this approach is a reasonable compromise, where the  majority of the disk is stable against disk self-gravity.
Additionally, the formation of pressure maxima stays within the stable inner regions of the disk, which are crucial for trapping and accreting particles.

\subsection{Including Dead Zone Edges}

As the viscosity is parameterized by assuming $\alpha$-prescription of \citep{ShakuraandSunyaev1973}, in order to explore the effect of viscosity on dust drift and dust accumulation, a set of $\alpha$ parameters were investigated. 
This includes a range of $\alpha = [0.01, 0.001, 0.0001]$. 
Additionally, the $\alpha_\mathrm{mod}$ parameter, which modifies the turbulent viscosity within the dead zone, varied in the range of $\alpha_\mathrm{mod} = [0.01, 0.001, 0.0001]$.

The dead zone edges are defined with an initial configuration of the $r_\mathrm{dze,i}$ and $r_\mathrm{dze,o}$ parameters.
The inner edge of the dead zone was set to the snow line of the disk, that lies at a distance of $2.7 $ AU from a typical solar type star \citep{Hayashi1981}.
Furthermore, according to \cite{MatsumuraandPudritz2005}, the outer edge of the dead zone was set to $r_\mathrm{dze,o} = 24$ AU.

Previous 2D hydrodynamical studies have shown that the properties of this viscosity transition significantly influence the characteristics of resulting 2D structures, such as vortex lifetime, azimuthal elongation, and strength \cite[e.g.,][]{Tarczayetal2022}.
Thus, even in 1D models where explicit vortex formation cannot be investigated, understanding the impact of the transition's width on the resulting pressure maxima is crucial. 
Therefore, the parameter space of $\Delta r_\mathrm{dze,i/o}$ covers a range of [1.0, 1.5, 2.0] in these models.

\begin{figure*}
    \centering
    \includegraphics[width=\linewidth]{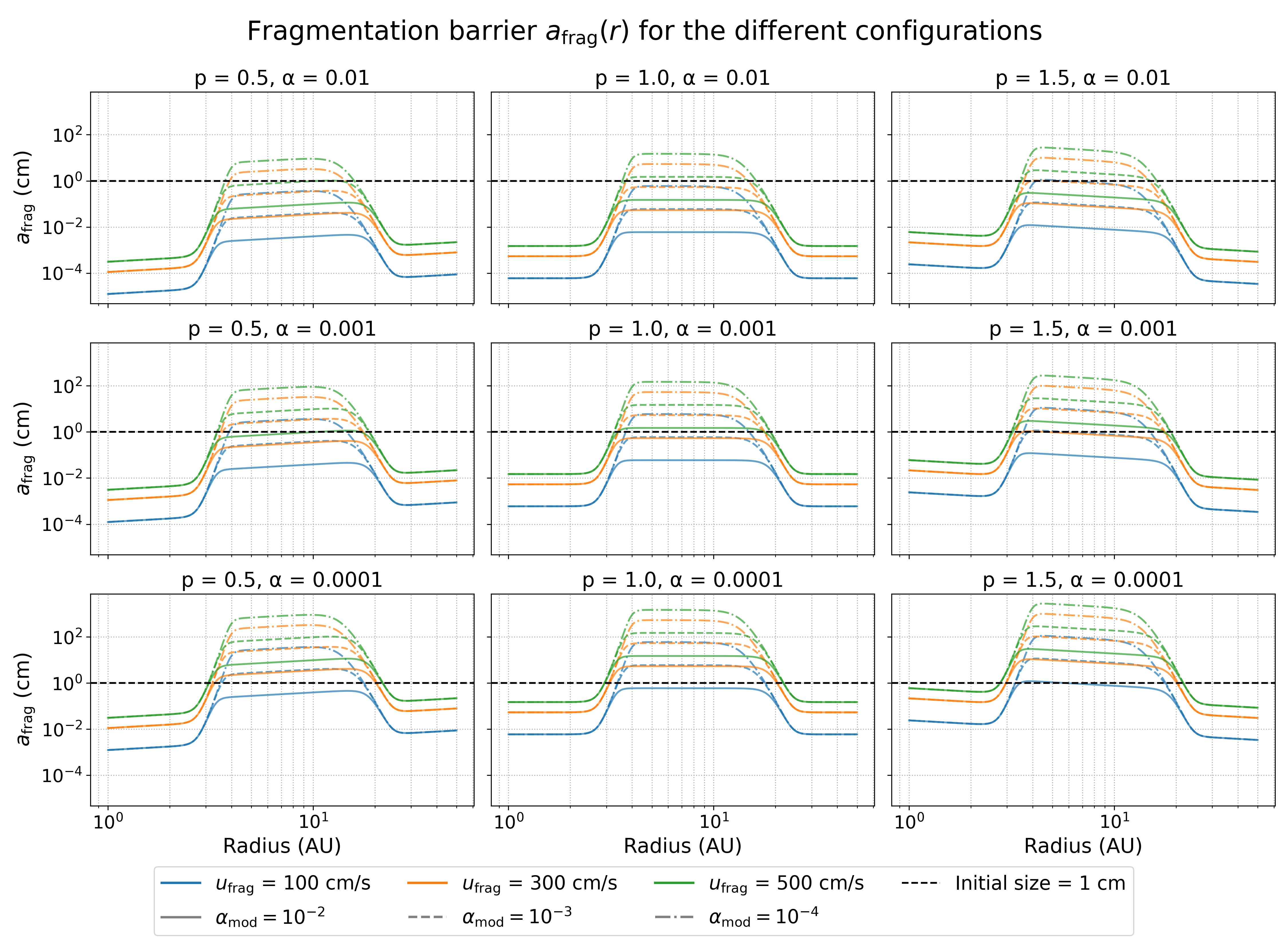}
    \caption{The radial distribution of the fragmentation barrier, $a_\mathrm{frag}(r)$, for different disk configurations. 
    Each panel corresponds to a combination of the surface density exponent $p$ and the global $\alpha$ viscosity parameter. 
    Curves are color-coded by the critical fragmentation velocity $u_\mathrm{frag}$ (blue: 100 cm/s, orange: 300 cm/s, purple: 500 cm/s), and line style indicates the depth of the viscosity reduction $\alpha_\mathrm{mod}$ (solid: $10^{-2}$, dashed: $10^{-3}$, dash-dotted: $10^{-4}$). 
    The horizontal dashed line marks the applied initial dust size of 1 cm.
}
    \label{fig:init_size_distr}
\end{figure*}
 
\subsection{Dust Particles}

For the sake of simplicity, the dust component is represented by $N_\mathrm{p} = 1000$ representative particles,matching the number of radial grid cells ($N_\mathrm{grid}$).  
This choice allows for a direct comparison of dust properties within the corresponding (initial) grid cell of the gas disk. 
With a lower number of particles, the representative dust particles would need to cover a larger radial range than the initial grid cell. 
This would lead to a loss of resolution for local gas disk properties (e.g., pressure maxima) and inaccurately represent the dust mass distribution within the given grid cells. 
Thus, this resolution is considered minimally sufficient for reliably tracking dust dynamics within the simulations, while maintaining computational efficiency.

The initial dust particle size is set to 1 cm in each cell throughout the whole simulation grid.
Note that this uniform initial size may conflict with the local maximum size allowed by the \cite{Birnstieletal2012} barriers, in certain disk configurations, i.e., $\alpha=0.01$, or $u_\mathrm{frag} = 100$ cm/s cases, see Figure\,\ref{fig:init_size_distr}.
This is particularly relevant outside the dead zone, where the fragmentation barrier $a_\mathrm{frag}$ is especially low due to the combination of high viscosity and low fragmentation threshold velocity.
However, the model implicitly handles such discrepancies as particle sizes are immediately adjusted to the physically permitted limit during the first timestep.

The fragmentation threshold velocity, $u_\mathrm{frag}$, is a crucial parameter defining the collision energy required for particle fragmentation. 
This value is set in the simulations, specifically $u_\mathrm{frag} = [100, 300, 500]$ cm/s, consistent with experimental measurements for silicate grains and theoretical considerations as discussed in Section \ref{sec:frag_barrier}. 
The intrinsic density of each particles ($\rho_\mathrm{p}$) is set to a common silicate value of 1.6 g/cm$^3$ uniformly for the whole setup.
The initial dust-to-gas ratio ($\epsilon$) was set to 0.01 for the whole set of simulations.

\section{Results}

In the explored parameter space, in all cases, density and pressure maxima at the edges of the dead zone develop.
These regions act as effective pressure traps for the dust particles, and their properties, including their onset and evolution, are highly dependent on the physical parameters and geometry of the disk.

For clarity and conciseness, only a set of example figures is shown in this section.
The full set of figures spanning the entire parameter grid is available in the Supplementary Material.

The main findings are detailed in the following subsections.

\subsection{Geometric Aspects of the Disk}

\begin{figure}
    \centering
    \includegraphics[width=\linewidth]{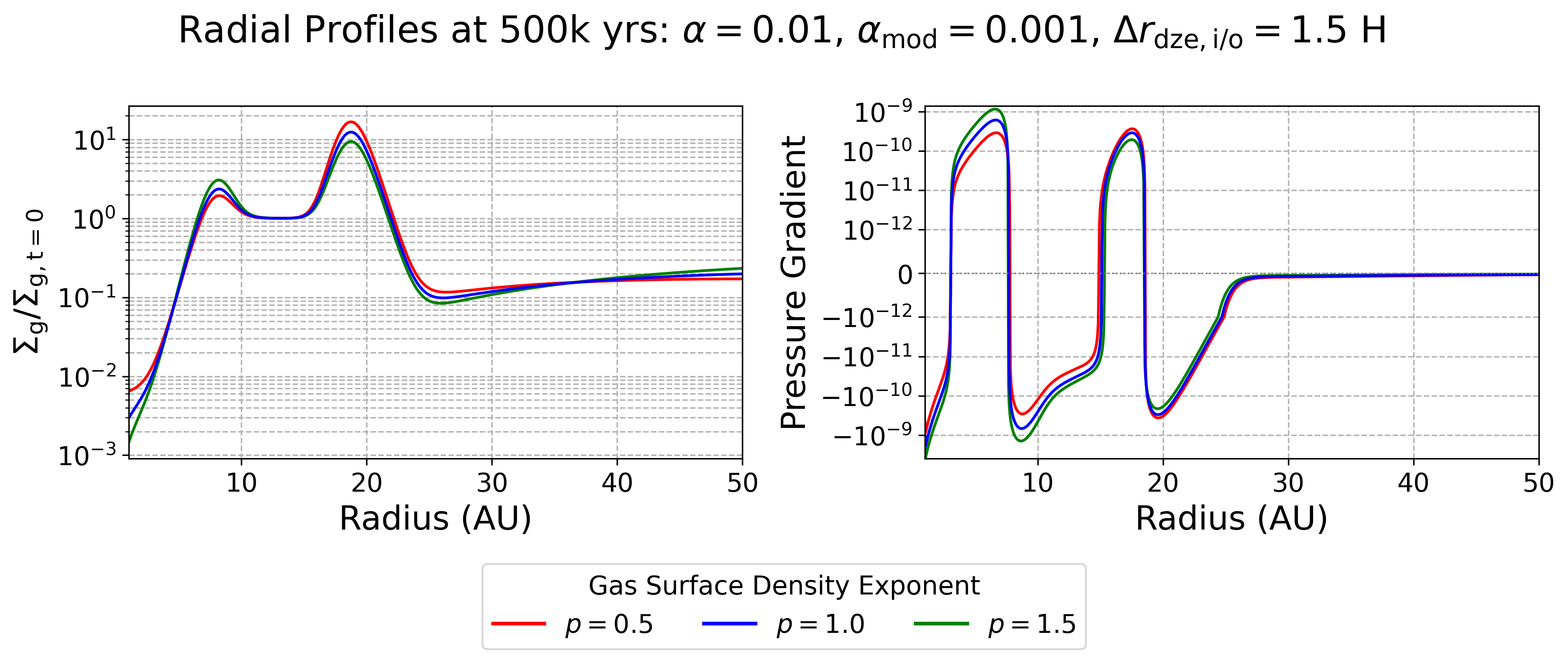}
    \caption{The normalized gas surface density profiles ($\Sigma_\mathrm{g}/\Sigma_\mathrm{g,t=0}$) and pressure gradient at $t=5\times10^5$ yrs for a given parameter set ($\alpha = 0.01$, $\alpha_\mathrm{mod} = 0.001$, $\Delta r_\mathrm{dze,i/o} = 1.5\,H$) for $p=0.5$ (red), $1.0$ (blue) and $1.5$ (green) disk profiles. 
    As the $p$ parameter increases, the difference between the inner and outer density maxima tends to equalize, resulting in a flatter overall profile.
    }
    \label{fig:1dprofile}
\end{figure}

\begin{figure}
    \centering
    \includegraphics[width=\linewidth]{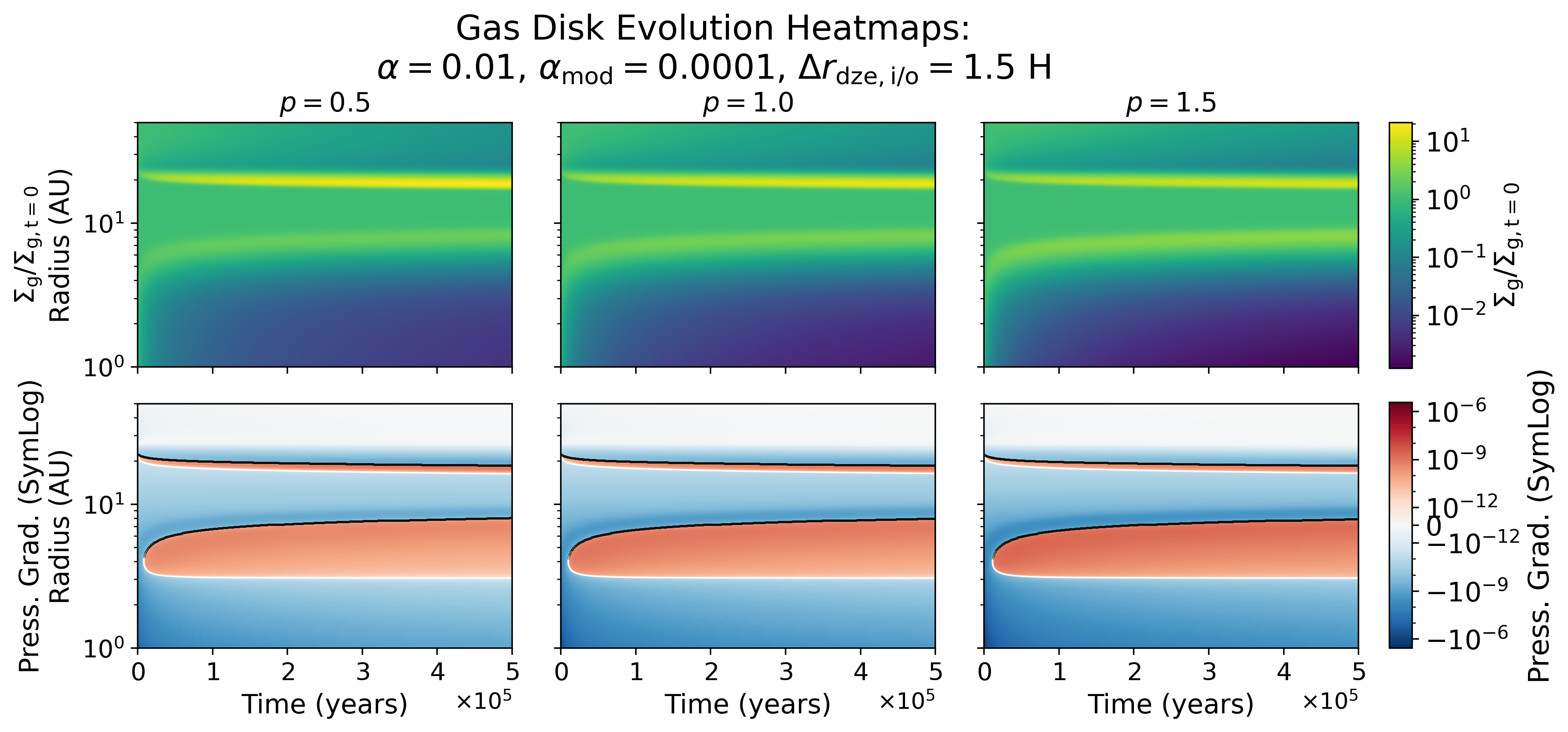}
    \caption{The evolution of normalized surface density ($\Sigma\mathrm{g}/\Sigma_\mathrm{g,t=0}$) of the gas and pressure gradient over time assuming a configuration of $\alpha =0.01$, $\alpha_\mathrm{mod} = 0.001$ and $\Delta r_\mathrm{dze,i/o} = 1.5$. Left panel corresponds to $p = 0.5$, while middle and right panels present $p=1.0$ and $p=1.5$ geometry, respectively.}
    \label{fig:heatmap}
\end{figure}

\begin{figure}
    \centering
    \includegraphics[width=\linewidth]{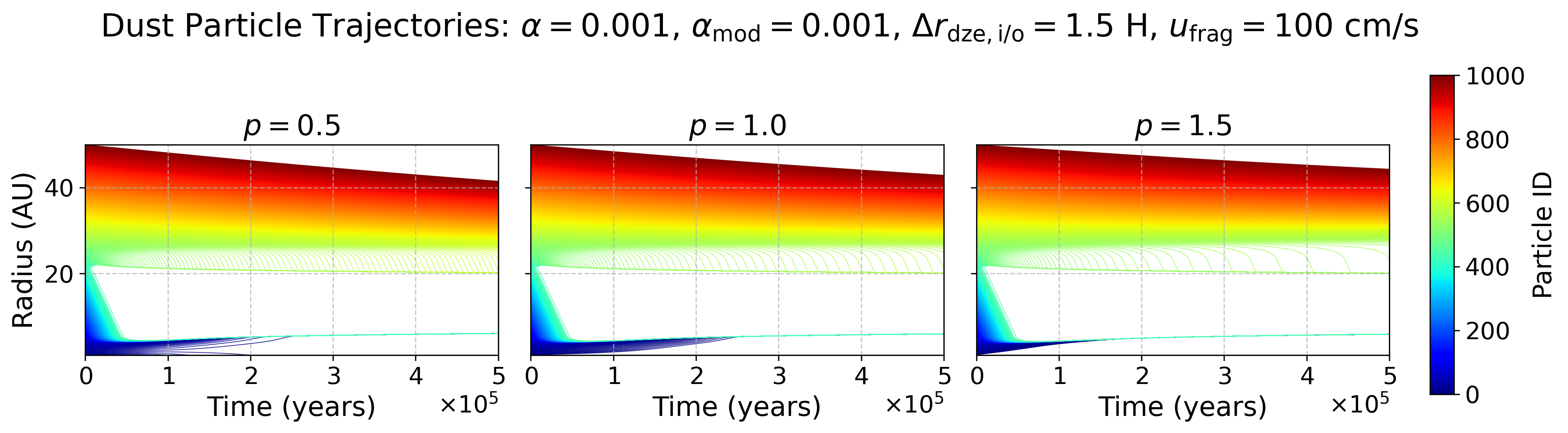}
    \caption{Individual dust particle trajectories assuming a disk configuration of $\alpha =0.001$, $\alpha_\mathrm{mod} = 0.001$ and $\Delta r_\mathrm{dze,i/o} = 1.5\,H$. Left panel corresponds to $p = 0.5$, while middle and right panels present $p=1.0$ and $p=1.5$ geometry, respectively. Fragmentation velocity $u_\mathrm{frag}$ is set to  $100$ cm/s. The trajectory of each particle is color-coded according to its unique identifier (ID) defined at the initial time
step.}
    \label{fig:part_trajectory}
\end{figure}

In this paper, different disk geometries were investigated, defined by the power-law index of the surface density profile ($\Sigma_\mathrm{g}\propto r^{p}$), see Section \ref{sec:simsetup} and \ref{tab:setup} for further details on disk setup.

Figure \ref{fig:1dprofile} shows the final ($t=1\times 10^5$ years), normalized ($\Sigma_\mathrm{g} /\Sigma_\mathrm{g,t=0}$) radial density profiles, where the gas surface density is divided by its initial value, so that all subsequent variations are expressed relative to the starting profile.
Assuming $\alpha = 0.01$, $\alpha_\mathrm{mod} = 0.001$ with a width of the viscosity transition region of $\Delta r_\mathrm{dze} = 1.5\,H$ at both the inner and outer edges of the dead zone.
Red, blue, and green lines demonstrate $p=0.5$, $1$ and $1.5$ disk geometries, respectively.
In Figure\,\ref{fig:1dprofile} it can be seen that as $p$ increases, the pressure gradient (right panel) and the normalized density contrast (left panel) at the inner edge of the dead zone both strengthen. 
In contrast, the outer pressure peak weakens with increasing $p$.
Thus, the radial $\Sigma_\mathrm{g}/\Sigma_\mathrm{g,t=0}$ profiles for each of the three cases show that increasing $p$, the difference between the inner and outer normalized density maxima tends to equalize, resulting in a less pronounced contrast between the two peaks . 
Concurrently, the inner region of the disk experiences a more significant depletion of gas relative to its initial state, whereas the outer regions maintain a gas density closer to their initial values.

The time evolution of the normalized gas surface density (and the corresponding pressure gradient) is presented in Figure \ref{fig:heatmap} for the same disk setup as presented in Figure \ref{fig:1dprofile}. 
In all panels of Figure\,\ref{fig:heatmap}, the x-axis represents time, the y-axis indicates the radial distance. In the upper panels the color coding illustrates the magnitude of variation of the gas surface density with respect to the initial one ($\Sigma_\mathrm{g} /\Sigma_\mathrm{g,t=0}$). 
The lower panels present the evolution of the pressure gradient for both the three disk geometries.
Utilizing a divergent color scheme, the positive and negative pressure gradient regimes can be easily distinguished.
Zero pressure gradient regions are indicated by white contours on each plots. 
Additionally, to highlight pressure maxima, which are known to effectively capture dust particles, black contours are used for these regions.

Consistent with Figure \ref{fig:1dprofile}, Figure \ref{fig:heatmap} also demonstrates the differences caused by varying $p$ parameters: as $p$ increases, the distinction between pressure maxima becomes less pronounced. 
Concurrently, the magnitude of the pressure gradient itself becomes less strong.
This trend is visible on a semi-logarithmic scale on the lower panels  in Figure\,\ref{fig:heatmap}.
A comprehensive set of similar heatmaps covering the entire parameter space is presented in the online Supplementary Figures S1/1 -- S1/9.
The trends observed in these figures suggest that increasing the $p$ parameter consistently leads to a decrease of the contrast in all cases, independent of the viscosity of the gas. 
For instance, assuming $\alpha_\mathrm{mod} = 0.0001$ and $\alpha = 0.01$, $\Sigma_\mathrm{g}/\Sigma_\mathrm{g,t=0} \approx 14$, $13$ and $10-11$ for $p=0.5$, $1.0$ and $1.5$, respectively.

Considering individual dust particle trajectories, it can be seen in Figure \ref{fig:part_trajectory} that as $p$ increases, the dust drift from the outer regions takes more time. 
This tendency becomes more pronounced for higher $p$ values and high viscosity models. 
Additionally, the feeding zone of the inner edge of the dead zone shrinks with increasing $p$.

\begin{figure}
    \centering
    \includegraphics[width=\linewidth]{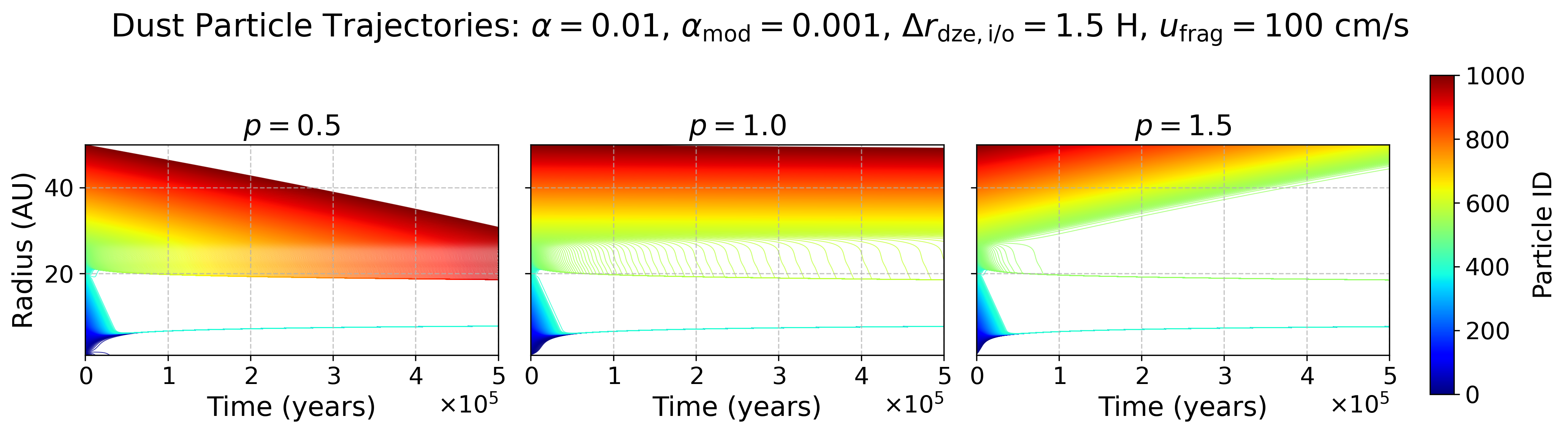}
    \caption{This figure presents the most pronounced effect of disk geometry in the case of $\alpha=0.01$, $u_\mathrm{frag} = 100$ cm/s models for $p=0.5$, $1.0$ and $1.5$ geometries, respectively. 
    It can be easily seen that in $p=1.5$ models, particles beyond the outer edge of the dead zone tend to drift outwards, escaping the computational domain within on the scale of  $\sim 5\times 10^{5}$ years.} 
    \label{fig:dustlossouter}
\end{figure}

In some cases, the initial amount of dust particles beyond the outer edge of the dead zone, may escape over time.
For example, in the case of a  configuration of $\alpha = 0.01$ and $p=1.5$, $u_\mathrm{frag} = 100$ cm/s, particles may eventually reach the outer boundary of the computational domain within $5\times 10^{5}$ years at 50\,AU, indicating a massive particle loss from the outer parts of the disk  (see Figure\,\ref{fig:dustlossouter}). 
A comprehensive view of individual dust trajectories for each parameter set can be found in the Supplementary data, see Figures S2/1 -- S2/27.

\subsection{The Effect of Viscosity}

\begin{figure}[!ht]
    \centering
    \includegraphics[width=\linewidth]{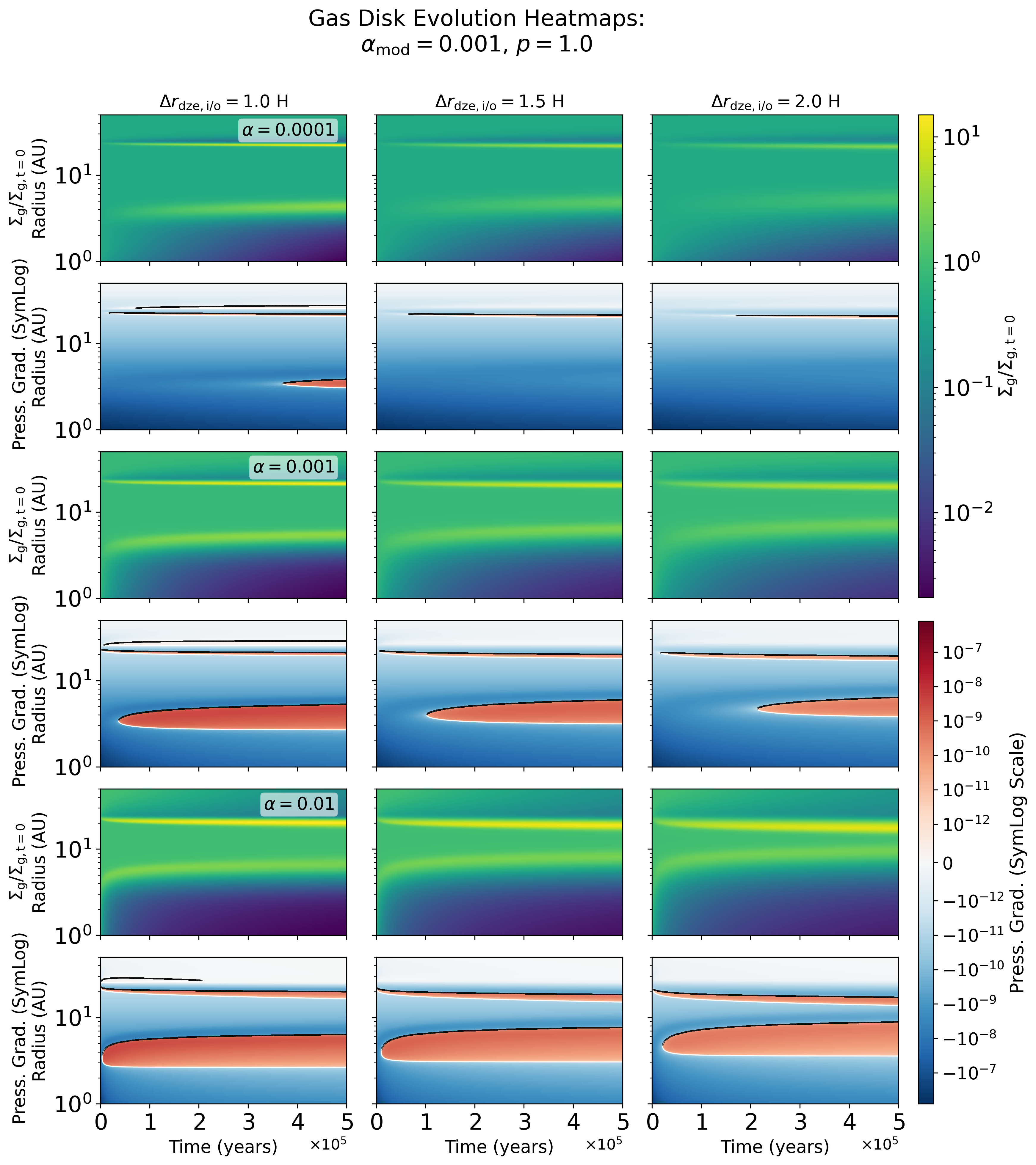}
    \caption{The effect of viscosity and the width of the viscosity reduction on the evolution of the gaseous material (presented by the normalized gas surface density, $\Sigma_\mathrm{g}/\Sigma_\mathrm{g,t=0}$ and the pressure gradient) in the disk assuming $\alpha_\mathrm{mod} = 0.001$ with a geometry of $p=1.0$. It can be seen that increasing $\alpha$ and decreasing $\Delta r_\mathrm{dze,i/o}$ leads to a sharper density contrast. Additionally, increasing $\Delta r_\mathrm{dze,i/o}$, the migration of the density maxima is more pronounced.}
    \label{fig:comparealpha}
\end{figure}

\begin{figure}
    \centering
    \includegraphics[width=\linewidth]{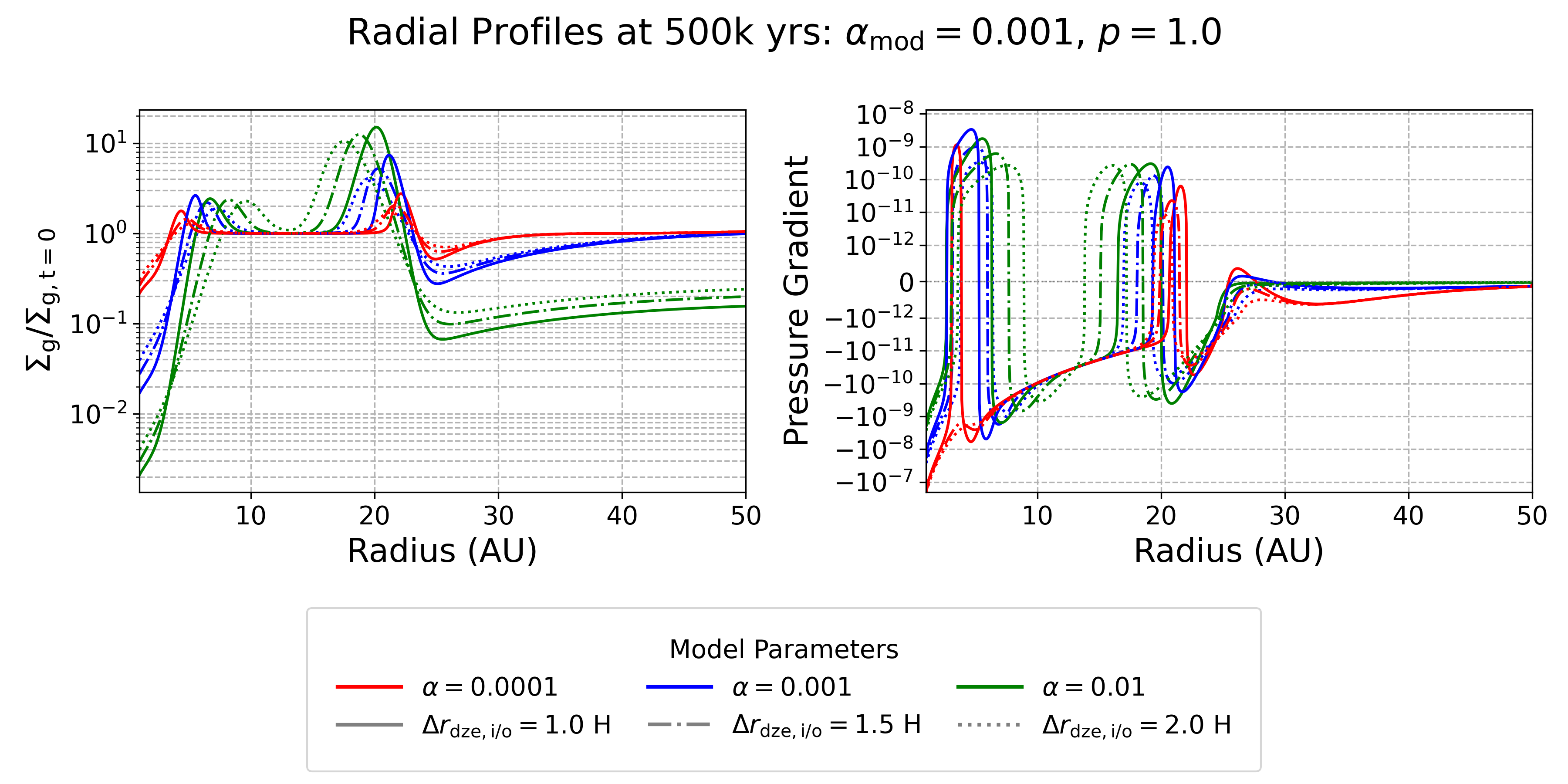}

    \caption{The normalized gas surface density profiles ($\Sigma_\mathrm{g}/\Sigma_\mathrm{t=0}$, left panel) and pressure gradient (right panel) at $t=5\times10^5$ yrs for a given parameter set assuming $\alpha_\mathrm{mod} = 0.01$ and $p=1.0$. Red, blue and green lines represent $\alpha=0.0001$, $0.001$ and $0.01$, while solid, dashed, dashed-dotted and dotted lines denote $\Delta r_\mathrm{dze, i/o} = 1.0$, $1.5$ and $2.0\,H$ cases, respectively. 
    }
    \label{fig:1d_sig_press_visc_compare}
\end{figure}

The results indicate that the evolution of the gaseous material is also highly dependent on the viscous properties of the gas.
It can be easily seen in Figure\,\ref{fig:heatmap}, that pressure maxima undergo radial movement over time.
Namely, the inner maximum migrates outwards and the outer maximum shifts inwards, potentially sweeping up dust particles within the dead zone.
Additionally, the results show that reducing the $\alpha$ viscosity parameter (from $0.01$ to $0.0001$), the migration of pressure maxima is less pronounced (see, e.g., Figure\,\ref{fig:comparealpha} and Figures S1 -- S3 in the Supplementary data).
In the case of low-viscosity models ($\alpha = 0.0001$), it can be seen, that the inner and outer edges of the dead zone remain closer to their initial radial positions. 
Concurrently, the contrast in gas surface density relative to the initial state becomes less distinct as $\alpha$ decreases.
Conversely, increasing the width of the viscosity transition region ($\Delta r_\mathrm{dze,i/o}$) from $1.0$ to $2.0\,H$ generally leads to a more pronounced migration of density maxima.
However, this also results in a delayed onset time for both pressure maxima. 

Furthermore, decreasing $\alpha$, the onset of both the inner and the outer pressure maximum is consistently delayed, or even inhibited within the simulation time. 
This trend is independent of $p$ and is true across all the examined parameter sets of models.
The effect is more pronounced for the inner maximum and is significant as it can lead to a rapid depletion of dust particles from the inner regions of the disk, inhibiting planetesimal formation. 
This effect can be seen in Figure\,\ref{fig:comparealpha}.
Specifically, for small $\alpha$ values and models with a larger transition zone (e.g., $\Delta r_\mathrm{dze,i/o} =1.5\,H$ and $2.0\,H$), the inner pressure maximum is effectively inhibited and does not form within the simulation's $10^5$-year timescale.
For example, the case of a sharp viscosity transition model ($\Delta r_\mathrm{dze,i/o}=1\,H$, the pressure maxima both at the inner and outer edges form rapidly (within a few $10^2 - 10^3$ yrs) in the high viscosity model ($\alpha$=0.01). 
On the contrary, decreasing $\alpha$ yields a slower pressure maxima formation: i.e., $\sim3\times10^4$, and $\sim3.5\times10^5$ years for $\alpha=0.001$, and $0.0001$, respectively.
Besides, increasing  $\Delta r_\mathrm{dze,i/0}$, in the case of $\alpha=0.0001$, the pressure maximum at the inner edge of the dead zone is not formed within the simulation time.

An additional, visible effect of viscosity on the evolution of gaseous material in the disk is due to the width of the viscosity transition region at the dead zone edges ($\Delta r_\mathrm{dze,i/o}$).
In the case of sharp viscosity transition ($\Delta r_\mathrm{dze,i/o} = 1.0\,H$), a secondary pressure maxima is formed beyond the outer edge of the dead zone in all cases (see, e.g., left panels of 
Figure\,\ref{fig:comparealpha}, and Figures S1/1 -- S1/9).
The formation and lifetime of the secondary maximum are clearly dependent on $\alpha$. 
For low viscosity models (e.g., $\alpha=0.0001$), it forms after $\sim 3\times10^4$ years and persists throughout the simulation. 
In contrast, for $\alpha=0.001$ and $0.01$, the formation is very rapid (on the magnitude of $\sim 1\times10^2-1\times10^3$ years). 
While it lasts throughout the simulation for $\alpha=0.001$, for $\alpha=0.01$ it decays after $\sim 1\times10^5$ years, compared to $\sim 2\times10^5$ years for $\alpha=0.001$ and $\sim 2.5\times10^5$ years for $\alpha=0.0001$.

\subsection{The Effect of Fragmentation}

In this paper, fragmentation is interpreted by the fragmentation velocity  (see more details in Section\,\ref{sec:frag_barrier} and Table\,\ref{tab:setup}). 
The results clearly show that increasing $u_\mathrm{frag}$ has a significant effect on the radial drift and thus on the overall evolution of the dust disk.
Increasing $u_\mathrm{drift}$ leads to a faster radial drift of particles towards the pressure maxima. 

\begin{figure}
    \centering
    \includegraphics[width=\linewidth]{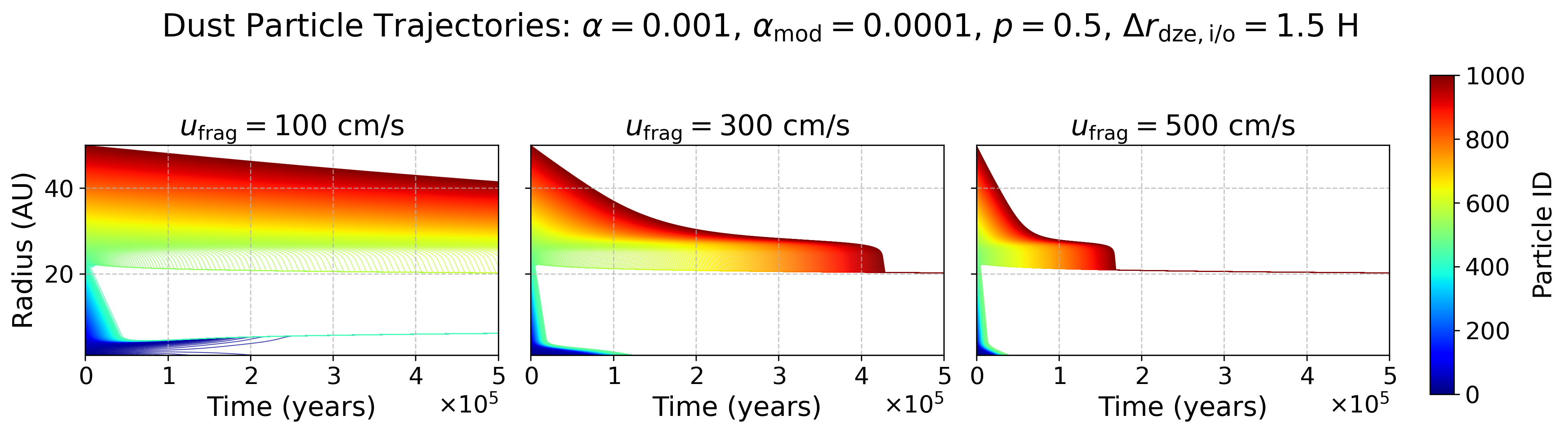}
    \includegraphics[width=\linewidth]{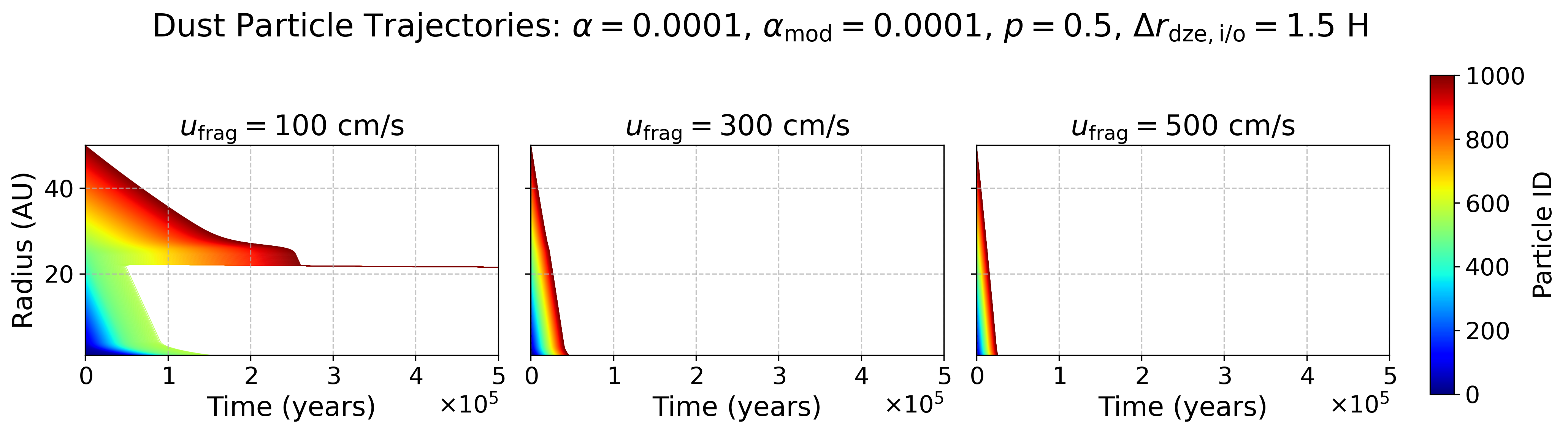}
    \caption{The effect of $u_\mathrm{frag}$ on individual dust trajectories in the case of $\alpha_\mathrm{mod}=0.0001$, $p=0.5$, $\Delta r_\mathrm{dze,i/o} =1.5\,H$. Upper and lower panels show $\alpha=0.001$ and $\alpha = 0.0001$ models, respectively.}
    \label{fig:compareufrag}
\end{figure}

However, this increased drift velocity can also lead to a more rapid depletion of dust from the entire disk, particularly in models where the inner pressure maximum forms on the magnitude of $1\times10^5$ years (i.e., low $\alpha$ models). 
Thus, the evolution of dust particles is highly sensitive to the interplay between particle growth and the timescale of pressure maximum formation. 
If the fragmentation velocity is high, particles can grow to sizes that experience rapid radial drift. Whether they are trapped in a dead zone edge then depends on the timing and efficiency of pressure trap formation.

This effect can be seen in Figure\,\ref{fig:compareufrag}. 
For instance, models with a low fragmentation velocity ($u_\mathrm{frag}=100$ cm/s), particles undergo a slower drift, allowing dust particles to grow larger. 
This results in a more rapid saturation of the inner dead zone edge with decreasing $\alpha_\mathrm{mod}$.
In contrast, for models with higher $u_\mathrm{frag}$ values (i.e., $300$ cm/s and $500$ cm/s), the particles drift inwards on a timescale that is faster relative to the formation of the pressure maxima. 
Thus, dust particles are depleted from the whole disk on a timescale on the order of $1\times10^3 - 1\times10^4$ years. 
This highlights a critical balance between particle growth and trapping required for successful planetesimal formation.

\begin{figure}
    \centering
    \includegraphics[width=\linewidth]{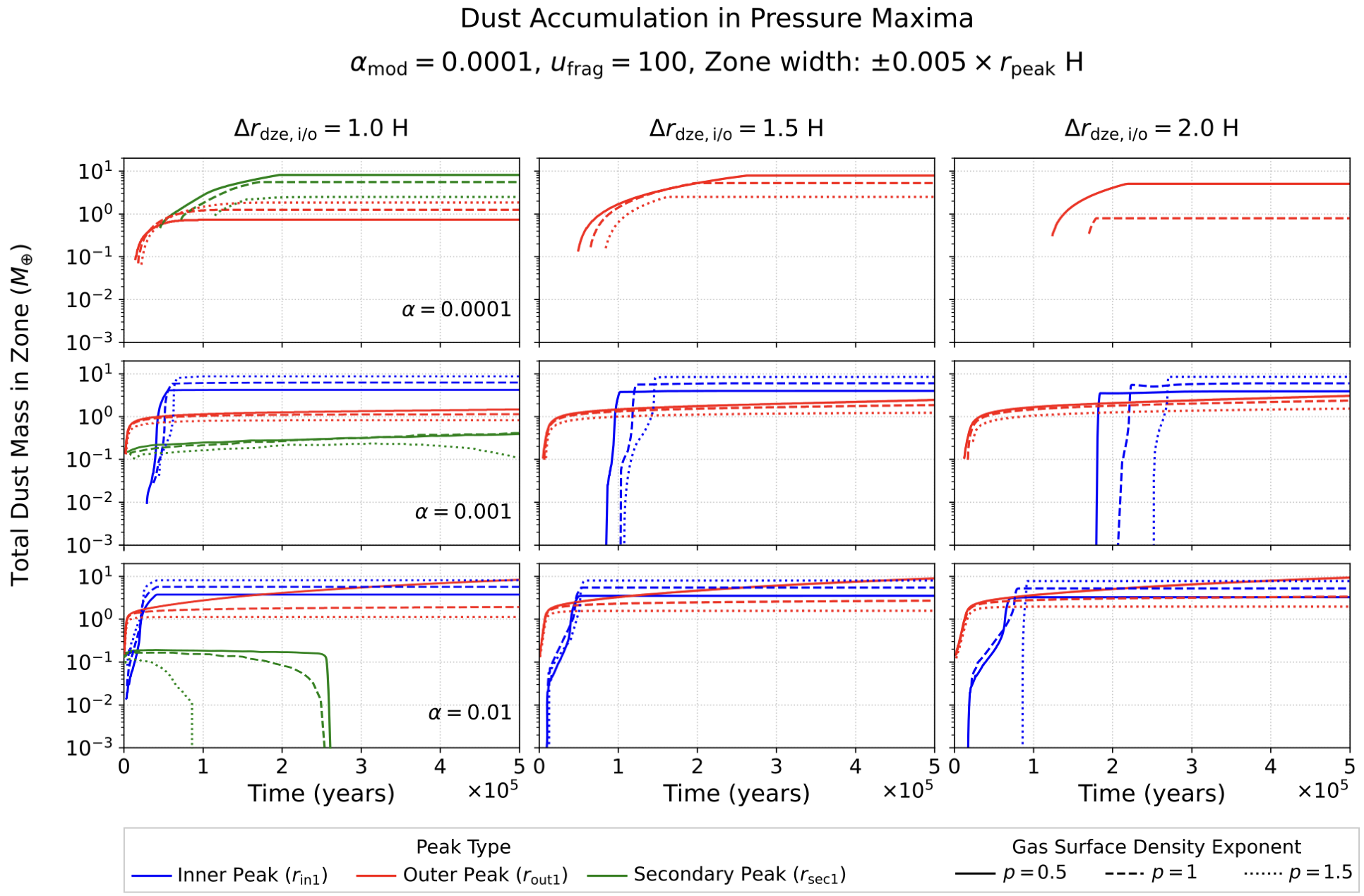}
    \caption{Mass accumulation in the case of $\alpha_\mathrm{mod} = 0.0001$, $u_\mathrm{frag} = 100$ cm/s models. The accumulated mass is integrated within a region of $\pm 0.005\,H$ around the corresponding pressure maximum. Blue, red, and green lines present the inner and outer edges of the dead zone, and the secondary maximum beyond the outer edge, respectively. $p=0.5$, $1.0$, and $1.5$ geometries are presented by solid, dashed, and dotted lines. $\alpha$ increases from top to bottom panels, while $\Delta r_\mathrm{dze,i/o}$ widens from left to right.}
    \label{fig:massaccum}
\end{figure}

\begin{figure}
    \centering
    \includegraphics[width=\linewidth]{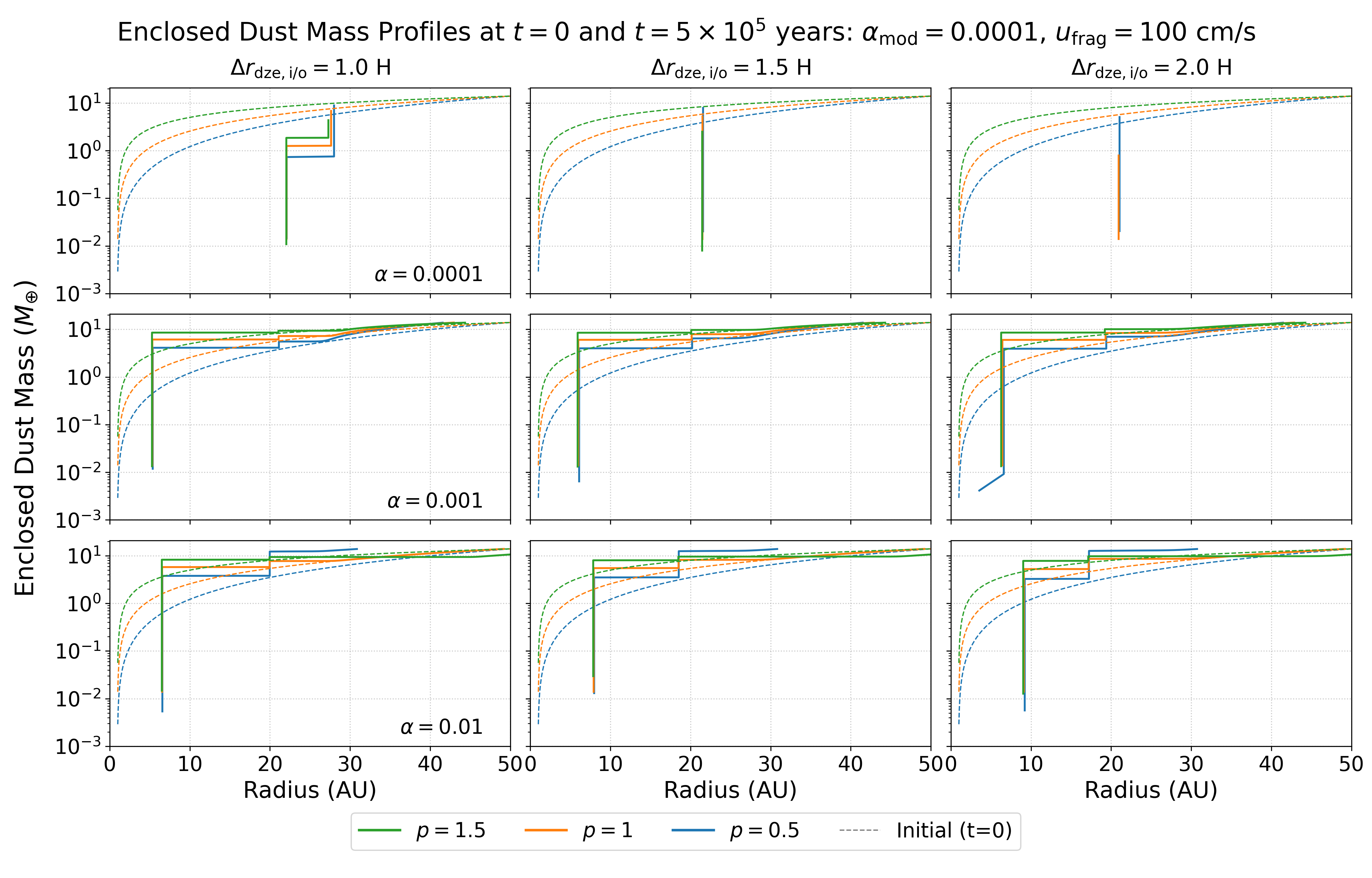}
    \caption{Enclosed dust mass profiles corresponding to the same disk models shown in the previous figure. Each panel represents a distinct combination of $\alpha$ and $\Delta r_\mathrm{dze}$, with colors indicating the surface density slope $p$. Dashed lines show the initial state ($t = 0$), while solid lines represent the final state at $t = 5 \times 10^5$ years. The enclosed mass is computed by integrating the dust surface density radially outward, providing a global measure of dust redistribution and accumulation across the disk.}

    \label{fig:encloseddust}
\end{figure}

\subsection{Mass Accumulation at the Dead Zone Edges}

A significant advantage of the RAPID code is its ability to monitor the evolution of individual particles separately. 
Since each particle is assigned to an initial representative mass, it is straightforward to measure the total mass of dust accumulating in pressure maxima.
As discussed in the previous sections, the evolution of the gaseous and dusty material is highly dependent on the physical properties of the disk and the dust particles themselves.
While in some cases, an additional secondary pressure maximum forms, in other configurations, the inner pressure maximum is not even formed within the simulation time.

Figure\,\ref{fig:massaccum} presents a set of  $\alpha_\mathrm{mod} = 0.0001$, $u_\mathrm{frag} = 100$ cm/s models.
The Figure shows that in the case of a sharp viscosity transition ($\Delta r_\mathrm{dze,i/o} = 1.0\,H$), the secondary pressure maximum is a highly effective dust trapping region. 
For instance, models with $\alpha=0.0001$, in the secondary maximum, a significant amount of dust can be accumulated, ranging from $6-8\,M_\oplus$, which is also dependent on the disk's geometry ($p$).
Furthermore, in the case of $\Delta r_\mathrm{dze,i/o} = 1.5$, and $2.0\,H$, in the absence of a secondary pressure maximum, the outer pressure maximum traps the dust. 
This can still be a substantial amount of dust, up to even $9-10\,M_\oplus$ for $p=0.5$, $\alpha=0.01$ models.

Figures S3/1--9 show a comprehensive view of mass accumulation governed by the physical properties of the simulation parameters.
The Figures clearly show that increasing $\alpha$ leads to a greater amount of dust being trapped at the inner dead zone edge. 
In general, increasing $\Delta r_\mathrm{dze,i/o}$ slightly decreases the amount of accumulated dust in pressure maxima. 
For instance, assuming a configuration of $\alpha = 0.01$, $\alpha_\mathrm{mod} = 0.0001$, $p=1.5$, $\sim 8.2\,M_\oplus$ -- $7.8\,M_\oplus$ from $\Delta r_\mathrm{dze,i/o} = 1.0$ to $2.0\,H$ at the inner edge of the dead zone (see Figure\,\ref{fig:compareufrag}).

Overall, the inner pressure maximum traps a greater amount of dust as the pressure power-law exponent ($p$) increases. 
For instance, in the case of $\alpha=0.01$ and $\Delta r_\mathrm{dze,i/o}=1.0\,H$, assuming a model with $p=0.5$, the mass of the piled up dust is $\sim 9\,M_\oplus$ at the outer pressure maximum and $\sim 3.5\,M_\oplus$ at the inner pressure maximum. 
In contrast, for a model with $p=1.0$, the mass at the inner pressure maximum increases to almost $6\,M_\oplus$, while the outer pressure maximum traps less than $2\,M_\oplus$. 
Further increasing $p$ to $1.5$, the inner pressure maximum traps over $8\,M_\oplus$, while the outer one traps only about $1\,M_\oplus$. 
In these models, the secondary pressure maximum consistently traps a small amount of dust on the magnitude of $0.1\,M_\oplus$, and this dust is typically lost from the region before the end of the simulation (i.e., trapped at the outer edge of the dead zone). 
This global redistribution is also reflected in the enclosed dust mass profiles (see Figure~\ref{fig:encloseddust}, and Figures S4/1 -- S4/9 in the Supplementary data).

\section{Discussion and the Concluding Remarks}

In the era of advanced 2D and 3D simulations, particularly with the rise of GPU-accelerated programming, the hydrodynamic simulations of protoplanetary disks have been revolutionized. These detailed, high-resolution models have enabled the investigation of large-scale, complex phenomena, such as RWI-induced vortices on a 2D or 3D grid. These advanced models, however, require a high computational cost, often requiring the use of powerful GPU clusters, large amounts of memory, and extended CPU time even on modern supercomputers.

Given these resource limitations, 1D models are still a useful tool for describing dust and gas interactions in protoplanetary disks, despite their simplified nature. Their primary advantages are the low computational resource requirements, which make them ideal for, e.g., exploring large parameter spaces and for a quick initial testing for 2D and 3D simulations. Furthermore, the optional straightforward CPU parallelization of these models facilitates a high number of simulation runs, which is beneficial for a large statistical grid of runs.

\subsection{Pressure Trap Formation and Evolution}

In all configurations, pressure maxima at the dead zone edges are formed. Although it is visible that the onset of pressure traps and the contrast of surface density ($\Sigma_\mathrm{g}/\Sigma_\mathrm{t=0}$) are highly dependent on the different physical parameters and geometry of the disk, such as viscosity, geometry, the width of the viscosity transition, etc.

\subsubsection{Viscosity and Viscous Timescale}

\begin{figure}
    \centering
    \includegraphics[width=0.75\linewidth]{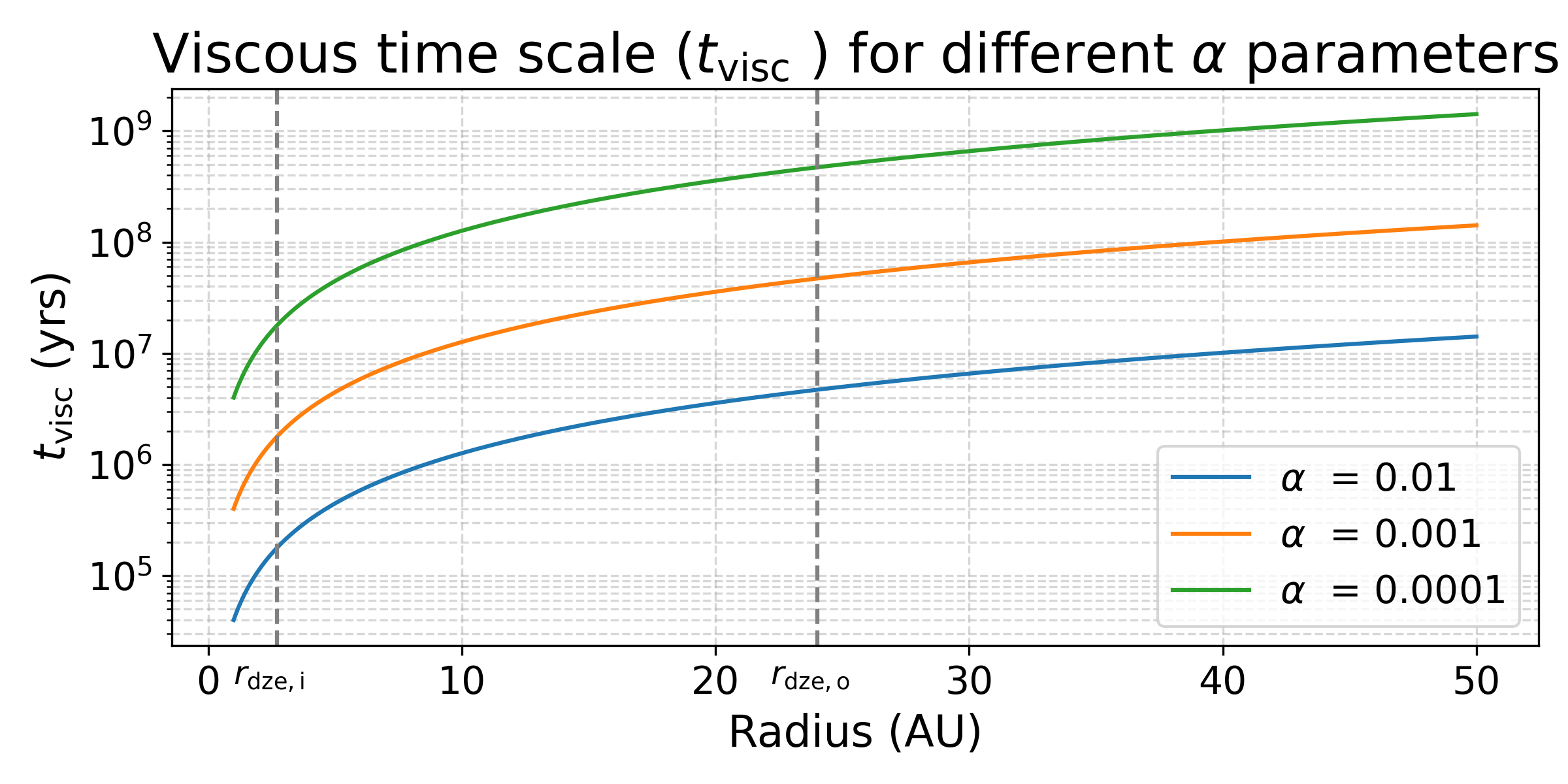}
    \caption{The viscous timescale ($t_\mathrm{visc}$) as a function of the $\alpha$-parameter applied in this paper.}
    \label{fig:tvisc}
\end{figure}

The \cite{ShakuraandSunyaev1973} $\alpha$-parameter (and thus the viscosity of the gas) strongly influences the formation and evolution of pressure maxima. 
This behavior can be understood in terms of the viscous timescale $t_\mathrm{visc}$ \citep[see, e.g.,][]{Pringle1981,Armitage2010,Armitage2011}, which governs how quickly pressure maxima can evolve:

\begin{equation}
    t_{\rm visc} \sim \frac{r^2}{\nu}.
\end{equation}

\noindent Assuming the $\alpha$-prescription (see Equation\,\ref{eq:viscosity}), this becomes:

\begin{equation}
    t_{\rm visc} \sim \frac{1}{\alpha H^2 \Omega_\mathrm{K}}.
\end{equation}

\noindent For typical disk parameters ($h = 0.05$, $\alpha = 10^{-2}$ to $10^{-4}$), the viscous timescale at the inner dead zone edge (2.7\,AU) ranges from $\sim 2 \times 10^5$ to $\sim 2 \times 10^7$ years, while at the outer edge (24\,AU) it ranges from $\sim 4 \times 10^6$ to $\sim 4 \times 10^8$ years (Figure\,\ref{fig:tvisc}). This yields a difference in $t_\mathrm{visc}$ between the edges of slightly more than one order of magnitude, meaning that the inner trap develops significantly earlier than the outer one.

As a result, reducing $\alpha$ delays the onset of both pressure maxima, and in extreme cases, may inhibit the formation of the inner trap entirely during the simulation time ($t = 5\times10^5$ years). This leads to rapid depletion of dust from the inner disk before a pressure trap can be formed.

These findings are consistent with previous 2D hydrodynamic simulations. For example, \cite{Tarczayetal2022} showed that RWI-induced anticyclonic vortices persist longer in high-viscosity disks, although they tend to be weaker than in low-viscosity counterparts.

\subsubsection{Disk Geometry and the Drift Velocity}
\label{sec:discussion_diskgeom}

\begin{figure}
    \centering
    \includegraphics[width=0.75\linewidth]{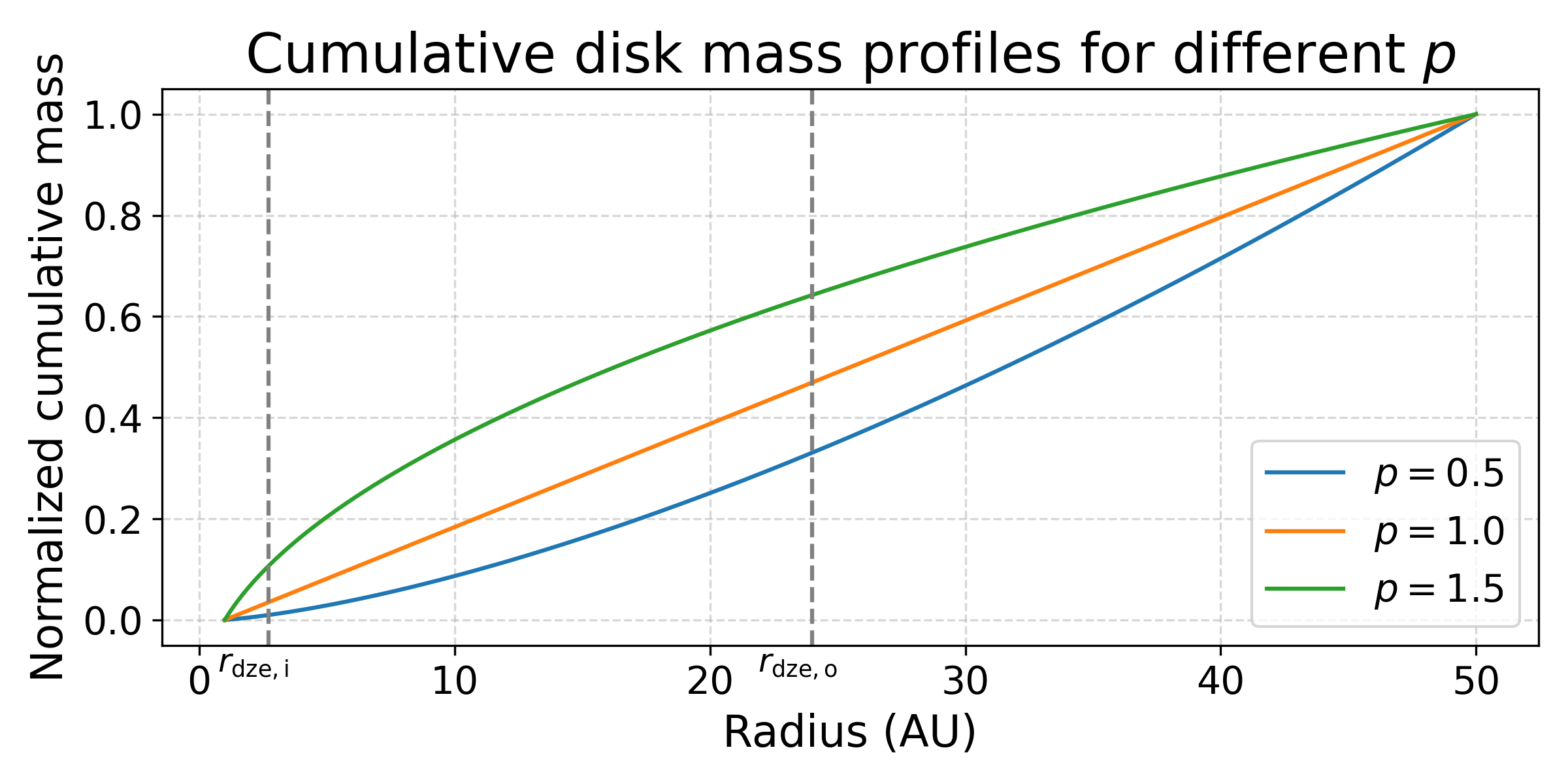}
    \caption{The normalized cumulative mass of the disk within the computational domain as a function of disk geometry ($p$).}
    \label{fig:mass_dist}
\end{figure}

The geometry of the disk has a major impact on the properties of the pressure traps. As shown in Figures~\ref{fig:1dprofile} and~\ref{fig:heatmap}, as the surface density power-law index ($p$) increases, the inner pressure maximum strengthens while the outer one weakens. This indicates that the geometry of the disk may determine which region becomes the primary site for dust accumulation and planetesimal formation.

This behavior arises from the interplay between the initial radial mass distribution of the gas and the dust, and the local pressure gradient of the gas that governs the drift velocity of the dust particles. Both are controlled by the $p$ parameter.
Namely, as $p$ increases, the surface density profile steepens, concentrating more mass toward the inner regions of the disk and reducing the available material in the outer parts, see Figure\,\ref{fig:mass_dist}. This enhances the inner pressure maximum to collect a huge amount of dust more efficiently with increasing $p$.
In contrast, the outer pressure maximum shows the opposite trend: increasing $p$ results in less dust being trapped in that region.

Moreover, in a locally isothermal disk, the radial drift velocity of particles is governed by the pressure gradient, and scales as
\begin{equation}
\label{eq:udriftsim}
    u_\mathrm{drift} \sim \frac{d\ln P}{d \ln r},
\end{equation}

\noindent see Equation\,\ref{eq:udrift}. 
\revv{As the disk evolves on the viscous timescale and develops pressure maxima at the dead zone edges, the pressure gradient that governs the drift becomes a local quantity that changes with radius and time, rather than a single global slope set by the initial surface density profile. Namely,}
as $p$ increases, the pressure profile steepens in the inner disk but flattens in the outer regions, leading to a weaker pressure gradient there. As a result, particles originating from the outer disk drift inwards more slowly. 
This can lead to cases where the outer disk is not fully depleted within the simulation time.
This is in agreement with what was described earlier by, e.g., \cite{Birnstieletal2012}.

\subsubsection{Migration of Pressure Traps}

The results showed that pressure maxima undergo radial drift over time. Namely, the inner maximum migrates outward while the outer maximum shifts inward. This drift is also influenced by the viscosity parameter $\alpha$ and the width of the viscosity transition ($\Delta r_\mathrm{dze,i/o}$): lower $\alpha$ and $\Delta r_\mathrm{dze,i/o}$ values tend to slow down the radial migration of the pressure maxima, while higher $\alpha$ and $\Delta r_\mathrm{dze,i/o}$ accelerates their drift. As a result, in low-viscosity disks, pressure traps remain closer to their initial locations for longer periods, potentially enhancing dust retention. In contrast, high-viscosity environments may cause the traps to migrate more rapidly, reducing their efficiency in capturing dust. This behavior likely arises because higher viscosity enhances angular momentum transport, accelerating the redistribution of gas and thus the drift of pressure maxima.

\subsubsection{Development of the Secondary Pressure Maximum}

In models with sufficiently sharp viscosity transitions (i.e., $\Delta r_\mathrm{dze,i/o} = 1\,H$), a secondary pressure maximum consistently forms beyond the outer edge of the dead zone, independent of $p$, $\alpha$, and $\alpha_\mathrm{mod}$. Although the onset and lifetime are sensitive to $\alpha$: in the case of $\alpha = 0.01$, it decays within $1\times10^5$–$2 \times 10^5$ years, while in $\alpha = 0.001$–$0.0001$ models, it persists throughout the entire simulation time ($5\times 10^5$ years). 

\begin{figure}
    \centering
    \includegraphics[width=\linewidth]{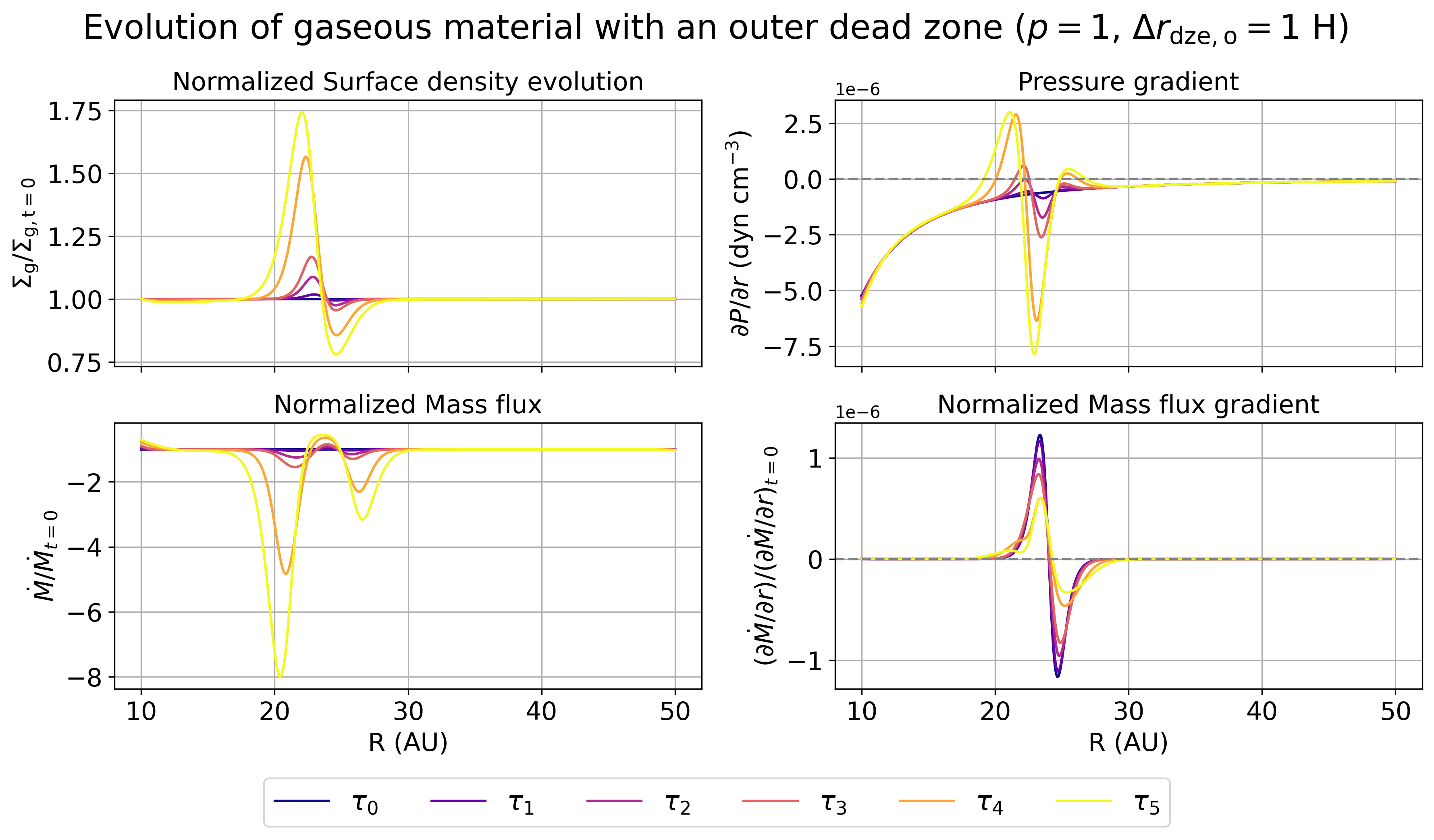}
    \caption{The evolution of the normalized surface density of the gas ($\Sigma_\mathrm{g}/\Sigma_\mathrm{g,t=0}$), the pressure gradient, the normalized mass flux ($\dot{M}/\dot{M}_{t=0}$), and the normalized derivative of the mass flux ($(\partial \dot{M}/\partial r) / (\partial \dot{M} /\partial r)_{t=0}$) at the vicinity of the outer edge of a dead zone ($p=1$, $\Delta r_\mathrm{dze,o} = 1\,H$)
    . Negative $\partial \dot{M} / \partial r$  indicates an enhanced inflow, leading to a dust accumulation. Snapshots are shown at successive simulation times, from $\tau_0$ to $\tau_5$, illustrating the temporal evolution of the disk.}
    \label{fig:modot}
\end{figure}

This secondary pressure maximum arises due to the sharp viscosity drop described at the outer edge of the dead zone. The resulting gradient in angular momentum transport leads to a local imbalance in the radial mass flux. The mass flux is defined as \citep[see, e.g.,][]{LyndenBellandPringle1974,Pringle1981}:

\begin{equation}
\label{eq:massflux}
\dot{M} = -2\pi r \, \Sigma_\mathrm{g} \, u_\mathrm{gas,r}.
\end{equation}

\noindent The direction of the mass flux is governed by the radial velocity of the gas, $u_\mathrm{gas,r}$ (see Equation\,\ref{eq:ugas}). Namely, inward movement ($u_\mathrm{gas,r}<0$) yields a positive mass flux, while outward motion ($u_\mathrm{gas,r} > 0$) corresponds to a negative one.

At the vicinity of the viscosity transition, the radial derivative $\partial \dot{M} / \partial r$ becomes negative, indicating that more gas is flowing inward than can be transported outward. This leads to gas a accumulation just beyond the dead zone edge. This forms a dynamically induced pressure bump, even if the associated increase in the gas surface density is low. 
This effect is visible in the bottom-right panel of Figure\,\ref{fig:modot}, which shows the evolution of normalized gas surface density (in a $p=1$, $\Delta r_\mathrm{dze,o} = 1H$ model), pressure gradient, normalized mass flux, and the normalized mass flux gradient in the vicinity of the outer dead zone edge. In all three subplots, the normalized quantities, surface density, mass flux, and mass flux gradient, are shown relative to their initial profiles.
It can be seen in Figure\,\ref{fig:modot} that beyond the outer edge of the dead zone, the location where $(\partial \dot{M} / \partial r)/(\partial \dot{M} / \partial r)_{t=0} < 0$ coincides with the secondary pressure maximum.

\subsubsection{Fragmentation Velocity}

Fragmentation velocity ($u_\mathrm{frag}$) is also a key parameter in the overall dust evolution. If this $u_\mathrm{frag}$ is too high (i.e., $300 - 500$ cm/s in low viscosity, $\alpha = 0.0001$, models), particles are cleared from the disk on a timescale faster than the formation of the pressure traps. This highlights a sensitive balance between dust sizes—driven by both coagulation and fragmentation—and the timescale of pressure maximum formation. This behavior is due to the fact that higher fragmentation velocities allow particles to grow rapidly to the characteristic size where the drift velocity peaks. This results in rapid inward migration—often on the magnitude of $10^3 - 10^4$ years—well before pressure traps can form and capture them.

\subsubsection{Dust Mass Collecting Efficiency in Pressure Traps}

Significant mass can be accumulated in pressure maxima. 
The simulations show that a substantial amount of dust, on the magnitude of several Earth masses ($3 - 10\,M_\oplus$), can be trapped at the dead zone edges, or at the secondary maximum, beyond the outer edge of the dead zone. 
The location of the most efficient trap is highly dependent on the disk's geometry. 
For instance, the inner maximum becomes more effective as the power-law exponent ($p$) increases, due to the steepening surface density profile and the resulting pressure gradient structure, as discussed in Section~\ref{sec:discussion_diskgeom}.

\subsection{Numerical Validation of the Advective Behavior}

\begin{figure}
    \centering
    \includegraphics[width=0.8\linewidth]{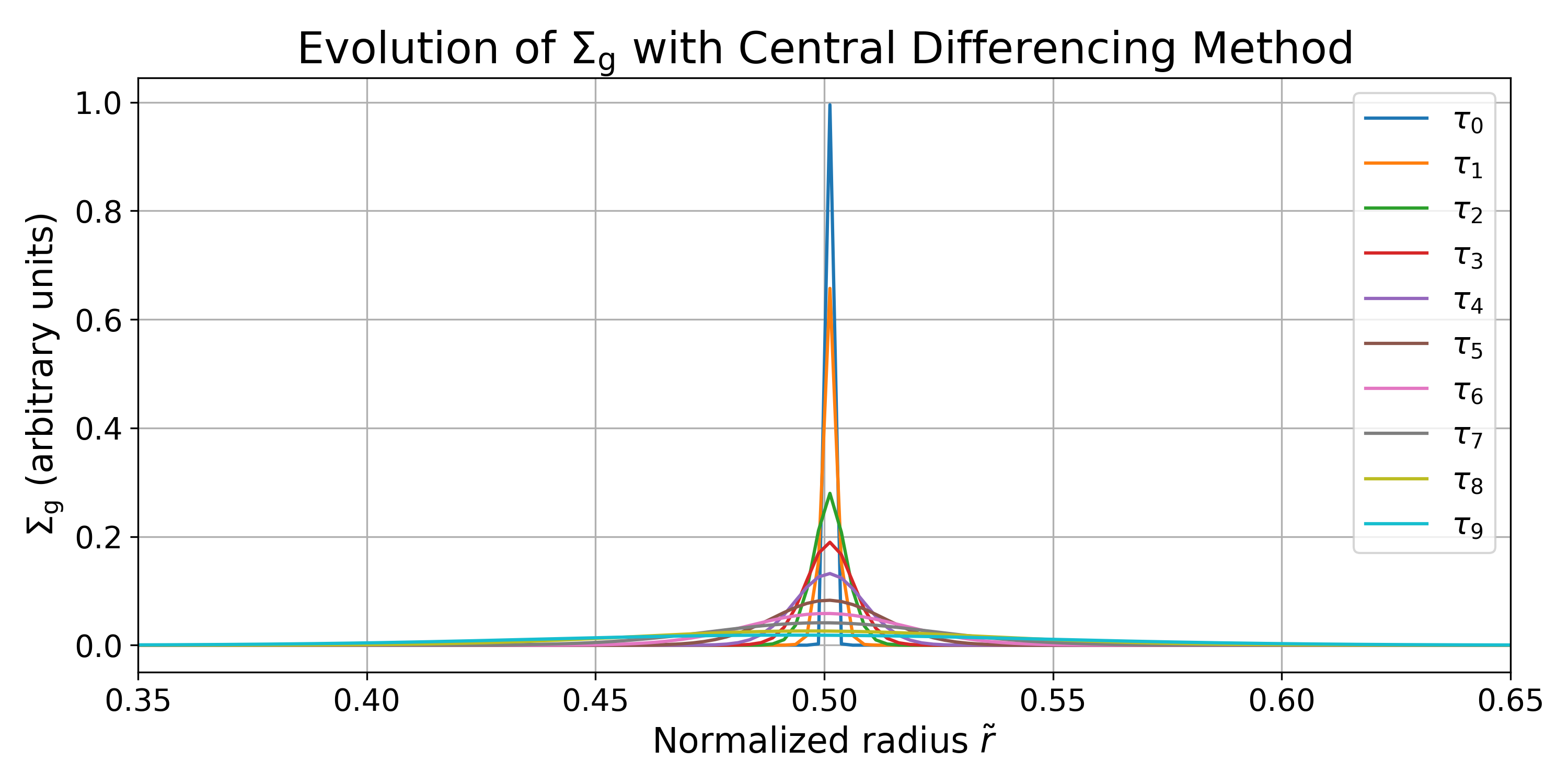}
    \caption{Time evolution of a localized surface density spike, shown as the numerical solution of Equation\,\ref{eq:gas_continuity_general}. The initial condition is a narrow Gaussian enhancement centered at $\tilde{r} = 0.5$. Over time, the profile broadens symmetrically due to diffusion and shifts slightly inward due to advection.}
    \label{fig:validation}
\end{figure}

As shown in Figure\,\ref{fig:dustlossouter}, under certain conditions (e.g., $\alpha = 0.01$, $u_\mathrm{frag} = 100$ cm/s, $\alpha_\mathrm{mod} = 0.001$, $\Delta r_\mathrm{dze,o} = 1.5H$, $p = 1.5$), particles can leave the system through the outer boundary of the integration domain.
To confirm that the outward transport observed in our models is a physical effect rather than a numerical artifact, a controlled test using a localized spike in surface density was performed (Figure~\ref{fig:validation}). 
It can be seen that over time, the spike spreads out due to diffusion and shows a slight inward shift of its peak, driven by advection. However, the diffusive broadening extends in both directions, and this symmetric spreading allows some material to reach and cross the outer boundary.

In the outer disk, where the surface density is low, the maximum particle size remains small. Initially, particles near the outer edge are typically in the $10^{-3}$–$10^{-4}$\,cm range in the above-mentioned configuration ($\alpha = 0.01$, $u_\mathrm{frag} = 100$ cm/s, $\alpha_\mathrm{mod} = 0.001$, $\Delta r_\mathrm{dze,o} = 1.5H$, $p = 1.5$), see Figure\,\ref{fig:init_size_distr}.
These grains are sufficiently small to remain tightly coupled to the gas, with typical Stokes numbers on the order of $10^{-5}$ or lower.
As a result, the outward flow of gas can thus sweep out well-coupled dust particles from the outer disk.

\subsection{Model Limitations and Caveats}

Although the RAPID is a 1D axisymmetric model, this study demonstrates the significant advantages in terms of computational efficiency and rapid parameter space exploration.
However, it also suffers from important limitations. 

\vskip 0.6em

1) Dust is modeled as discrete particles without continuous inflow from outer regions, meaning that the trapped dust mass represents an upper bound rather than a replenished quantity. 

\vskip 0.6em

2) Additionally, the disk mass was reduced to satisfy the Toomre $Q$ stability criterion (see Section\,\ref{sec:diskgeom}) and avoid the need to include self-gravity—an effect known to influence vortex dynamics and dust concentration in more massive disks \citep[e.g.,][]{RegalyandVorobyov2017}. 
Implementing self-gravity into future versions of the RAPID code would be essential for capturing these interactions accurately.

\vskip 0.6em

3) Furthermore, the current model does not include the backreaction of dust onto the gas.
 which can significantly alter pressure profiles and gas flow, especially in regions with high dust-to-gas ratios \cite[see, e.g.,][]{BenítezLlambay2018,Regalyetal2025}. 
 Including this feedback mechanism, along with self-gravity, in future 2D or 3D simulations will be crucial for a more complete understanding of disk evolution.

 \vskip 0.6em

 4) In \texttt{RAPID}, the dead zone is implemented via reduced viscosity with a smooth transition region. This transition is governed by a $\tanh$ function parameterized by $\alpha$ and $\alpha_\mathrm{mod}$. The applied prescription provides a fixed representation of the accretionally inactive region, rather than derived from the evolving disk structure. A more realistic formulation would link the viscosity transition to local physical quantities, such as surface density and temperature, allowing the bump to emerge self-consistently and respond to disk evolution.

 \vskip 0.6em

5) The current version of \texttt{RAPID} adopts a locally isothermal equation of state. Thus, the thermodynamics of disk evolution—such as heating from accretion or cooling—are not included.
 
 \vskip 0.6em

6) In addition to the above, it may also be important to implement the presence of a planetary embryo and the inclusion of disk wind in order to assess their impact on the gas and dust content of the disk, since it is well established that an embedded planetary embryo can open a gap in the disk, thereby influencing its subsequent evolution \citep[see, e.g., a recent study in 1D of][]{Sándoeretal2024}.

\backmatter

\section*{Supplementary information}

Supplementary figures covering the entire parameter grid are available in the online supplementary material. 
There are three supplementary files in total.

\section*{Acknowledgements}

I hereby dedicate this paper to Professor Bálint Érdi on his 80th birthday. 
He was one of my most respected professors at university, who introduced me to the beauty of celestial mechanics, although my research path turned in another direction. 
Beyond his expertise, he has always been a supportive mentor, one to whom I could turn to for guidance whenever needed.

This research was supported by the `SeismoLab' KKP-137523 \'Elvonal grant of the Hungarian Research, Development and Innovation Office (NKFIH), and by the LP2025-14/2025
Lendület grant of the Hungarian Academy of Sciences. 
The simulations were performed on the high-performance computing machine at HUN-REN CSFK CSI acquired through the EC Horizon2020 project OPTICON (Grant Agreement No. 730890).

This paper is written by the assistance of a large language model (e.g., Gemini by Google, GitHub Copilot and ChatGPT by OpenAI) for linguistic  refinement. 
All content and any errors remain the sole responsibility of the author.

\section*{Statements and Declarations}

\subsection*{Funding}
This research was supported by the `SeismoLab' KKP-137523 \'Elvonal grant of the Hungarian Research, Development and Innovation Office (NKFIH), and by the LP2025-14/2025
Lendület grant of the Hungarian Academy of Sciences. 
The simulations were performed on a high-performance computing machine acquired through the EC Horizon2020 project OPTICON (Grant Agreement No. 730890).

\subsection*{Competing interests}
The author declares that he has no competing interests.

\subsection*{Ethics approval and consent to participate}
Not applicable.

\subsection*{Consent for publication}
Not applicable.

\subsection*{Data availability}
The datasets generated and analyzed during the current study are available on Zenodo via DOI: 10.5281/zenodo.16579810.

\subsection*{Code availability}
The RAPID model code will be available at \url{https://github.com/tnehezd/RAPID/}.

\subsection*{Author contribution}
Single author: the author designed the model, conducted the analyzis, and wrote the manuscript.

\bigskip
\begin{flushleft}%
Editorial Policies for:

\bigskip\noindent
Springer journals and proceedings: \url{https://www.springer.com/gp/editorial-policies}

\bigskip\noindent
Nature Portfolio journals: \url{https://www.nature.com/nature-research/editorial-policies}

\bigskip\noindent
\textit{Scientific Reports}: \url{https://www.nature.com/srep/journal-policies/editorial-policies}

\bigskip\noindent
BMC journals: \url{https://www.biomedcentral.com/getpublished/editorial-policies}
\end{flushleft}



\bibliography{references} 

\end{document}